\documentclass{article}

\usepackage{amssymb,amsmath,amsfonts,amsthm}
\usepackage{graphicx}
\usepackage{ifpdf}
\usepackage{subfigure}
\usepackage{times}
\ifpdf\fi
\usepackage{url}
\usepackage{fullpage}

\newtheorem{theorem}{Theorem}
\newtheorem{lemma}[theorem]{Lemma}
\newtheorem{definition}{Definition}
\newtheorem{remark}{Remark}
\newtheorem{example}{Example}

\newcommand{\para}[1]{ \medskip \noindent {\bf #1}}
\newcommand{\eat}[1]{}
\newcommand{\bin}{ \operatorname{Bin}}
\renewcommand{\Pr}{\mathsf{Pr}}
\newcommand{\E}{\mathsf{E}}

\title{Differentially Private Publication of Sparse Data}

\author{Graham Cormode, Magda Procopiuc, Divesh Srivastava \\
AT\&T Labs--Research\\
\{graham,magda,divesh\}@research.att.com 
 \and Thanh T. L. Tran \\
 University of Massachusetts, Amherst \\
 ttran@cs.umass.edu}
\date{}

\begin{document}
\maketitle

\begin{abstract}
The problem of privately releasing data is to provide a version of a 
dataset without revealing sensitive information about the individuals 
who contribute to the data.  The model of differential privacy allows 
such private release while providing strong guarantees on the output. 
A basic mechanism achieves differential privacy by adding noise to the 
frequency counts in the contingency tables (or, a subset of the count 
data cube) derived from the dataset.  However, when the dataset is 
sparse in its underlying space, as is the case for most multi-attribute 
relations, then the effect of adding noise is to vastly increase the 
size of the published data: it implicitly creates a huge number of 
dummy data points to mask the true data, making it almost impossible
to work with. 

We present techniques to overcome this roadblock and allow efficient 
private release of sparse data, while maintaining the guarantees of 
differential privacy.  Our approach is to release a compact summary of 
the noisy data.  Generating the noisy data and then summarizing it 
would still be very costly, so we show how to shortcut this step, and 
instead directly generate the summary from the input data, without 
materializing the vast intermediate noisy data.  We instantiate this 
outline for a variety of sampling and filtering methods, and show how 
to use the resulting summary for approximate, private, query answering. 
Our experimental study shows that this is an effective, practical
solution, with comparable and occasionally improved utility over the 
costly materialization approach.
\end{abstract}

\section{Introduction}
A fundamental problem in data management is how to share real data
sets without compromising the privacy of the individuals who
contribute to the data. 
Great strides have been made on this problem in recent years, leading
to a growing science of ``anonymization''.
Such anonymization enables data owners to share data privately with
the public, external collaborators, or other groups within an
organization (the risk of private information leaking guides the
amount of masking applied to the data). 
Anonymization also allows businesses to retain detailed information
about their customers, while complying with data protection
and privacy legislation. 

Different techniques are relevant to different threat models: 
`syntactic' privacy definitions, such as $k$-anonymity and $l$-diversity, 
preserve more detail at the microdata level, but can
be susceptible to attack by determined adversaries~\cite{Kifer:09}. 
Differential privacy, introduced in a series of papers in the theoretical community,
is a more semantic definition. 
It has gained considerable traction due to its
precise privacy guarantees. See
recent tutorials on data anonymization for
further background and definitions~\cite{Chen:Kifer:Lefevre:Machanavajjhala:10,Cormode:Srivastava:09,Gehrke:Kifer:Machanvajjhala:10}.

The original model for differential privacy assumed an interactive
scenario: queries are posed to a ``gatekeeper,''  
who computes the true answer, adds random noise and returns the
result. 
The random noise is drawn from a  particular distribution, whose parameter is 
determined by the impact any individual can have on the answers to a
fixed number of queries. 
Once that number is reached, the gatekeeper stops answering. 
To avoid this restriction, we can release many
statistics about the data {\em en masse}. 
For example, release data in the form of contingency tables, with
appropriate noise added to each entry
\cite{Barak:Chaudhuri:Dwork:Kale:McSherry:Talwar:07}.  
This is equivalent to the result of various
groupby/count(*) operations on the original data (with noise).

\begin{example}
\label{eg1}
Consider a collection of geographic data, which records the commuting
patterns of a large population. This could be collected by a census bureau \cite{Machanavajjhala:Kifer:Abowd:Gehrke:Vilhuber:08}, or derived
by a cellphone provider based on calling patterns
\cite{Isaacman:Becker:Caceres:Koburov:Rowland:Varshavsky:10}. 
A contingency table, indexed by home 
and work locations, is a matrix: each cell contains the number of people who
commute from the given source to the given destination.  
We assume that the locations are sufficiently large areas so that
there are many cells with moderately large numbers in them. 
To release this data in a differentially private way, we add random
noise independently to each cell before publishing  
the matrix. The noise is relatively small (typically single
digit perturbations), so the large patterns in the data are preserved
while ensuring that no individual's contribution is revealed. 
\end{example}
\noindent {\bf Scalability Problem.} 
In theory, the above scheme is a useful approach to privately
release data. 
But a roadblock emerges, showing this method to be impractical. 
As we observe below, the contingency table can be orders of magnitude
larger than the original data. 
To satisfy privacy, we must add noise to {\em every cell}, including
those with zero count. 
Because the noise has a low probability of being zero, 
the resulting table is both large and very dense: we
cannot represent it compactly (such as by a list of non-zero cells).
The method is time and space consuming, to the extent where it becomes
unfeasible for many datasets. 

\begin{center}
\begin{table}[t]%
\centering
\begin{tabular}[t]{l|r|l}
 Dataset & Density $\rho$ & Source\\
 \hline
OnTheMap & 3-5\% & \url{lehd.did.census.gov} \\
Census Income  & 0.4-4\% & \url{www.ipums.org} \\
UCI Adult Data & 0.14\% & \url{archive.ics.uci.edu/ml} \\
Warehouse Data& 0.5-2\% & --- \\
\end{tabular}
\caption{Examples of datasets and their densities.}
\label{tblsparse}
\end{table}
\end{center}

Consider Example~\ref{eg1} for 
 a dataset of 10 million commuters, spread across 1 million
locations. 
The contingency table over home and work locations has $10^{12}$ entries. 
Just storing this table requires terabytes,
and generate $10^{12}$ random draws from the noise
distribution is extremely slow. 
We have gone from an original manageable data set with
$10^7$ rows to one that is $10^5$ times larger, making subsequent
analysis slow and unwieldy.
 
Table~\ref{tblsparse} shows several examples of widely-used datasets 
(further details of these datasets are explained in
Appendix~\ref{app:datasets}). 
For each dataset, we show the {\em density}, $\rho$, which is the
ratio between the original data size, and 
the size of the noisy contingency table. 
Hence, 
naively applying differential privacy generates output which is 
$1/\rho$ times larger than the data size. 

As Table~\ref{tblsparse} indicates, many natural datasets 
have low density in the single-digit percentage range, or less. 
This is the norm, rather than the exception: 
any data with several attributes $A_i$ of moderately high cardinality
$|A_i|$ leads to huge contingency matrices of size $\Pi_i |A_i|.$ 
For a table with $n$ rows, the density is at most 
$\rho = n/\prod_i |A_i|$. 
In the above examples, $1/\rho$ ranges from tens to thousands. 
This $1/\rho$ factor applies to the time required to prepare the private data,
and to store and process it. 
This is clearly not practical for today's large data sizes. 

\para{Our Contributions.} In this paper we propose new techniques for
making differential privacy scalable over low-density (sparse) datasets. 
Our algorithms compute privacy-preserving outputs whose size is
controlled by the data owner. 
Usually, this is chosen  close to the size of the original data,
or smaller.
Moreover, their running 
time is proportional to the size of the output, and independent of the
size of the contingency table.  
This is a crucial step towards making differential privacy
practical in database applications. 
 
As  discussed above, the anonymization process changes originally
sparse datasets into dense ones.  
To meet the strict differential privacy definition, there must be some 
probability of creating noise in {\em any} cell, else a powerful analyst
could eliminate some possibilities for a targeted individual. 
However, once the noisy contingency table is generated, we can safely
return only a random sample of it, instead of the entire data. 
We refer to this sample as the differentially private {\em summary}
of the data. 
Our approach relies on the well-known property
that any post-processing of differentially private data remains
differentially private;
see~\cite{Li:Hay:Rastogi:Miklau:McGregor:10,Kifer:Lin:10}.  

While post-process sampling alleviates the storage requirements of the
output, it does not help the overall computation cost. 
Hence, this approach is effective for the data recipient, 
but still very costly for the data owner. 
Instead, we propose a new technique: we show how to 
directly generate the summary from the original data, without
computing the huge contingency table first.  
This requires some care:  The (random) two-step process of generating
noisy intermediate 
data and then applying some sampling method over it implies a
probability distribution over possible summaries.  
Our goal is to provide one-step methods that create
summaries with {\em the same} distribution over possible
outputs, while being much more efficient. 
We design several one-step summary generation methods, which are
provably correct and run in time proportional to the output size.

Our algorithms depend on the specific sampling method employed. There
is a wealth of work on data reduction techniques. 
We focus on some of the best known methods, such as 
filtering, priority sampling and sketching~\cite{Garofalakis:Gehrke:Rastogi:02}.
They enable vast quantities of data to be reduced to much more
manageable sizes, while still allowing a variety of queries to be
answered accurately. To summarize:
\begin{list}{\labelitemi}{\leftmargin=1em}
\addtolength{\itemsep}{-1ex}
\item 
  We formalize the problem of releasing sparse data, and describe
  appropriate target summaries: filters, samples and sketches. 
\item
 We show techniques to efficiently generate these summaries, by
 drawing directly from an implicit distribution over
 summaries. 
\item 
 We perform an experimental study over a variety of data sets.
 We compare the utility of our summaries to that of the large intermediate tables,
  by running a large number of queries over both, and comparing their relative accuracies.
 We conclude that our significantly smaller summaries have similar utility to the large
 contingency tables, and in some cases, they even improve on it.
\end{list}
For background, a primer on ideas and prior work in 
differential privacy and data
summarization is presented in Appendix~\ref{app:background}. 

\section{Creating Anonymized Summaries}

\subsection{The Shortcut Approach}

Our goal is to efficiently release a summary of a dataset with privacy
guarantees. 
Formally, let $M$ be the original data, represented as a
(potentially huge) contingency table. 
We assume that $M$ has low density, i.e.,
the number of non-zero entries in $M$, denoted by $n$, is 
much smaller than the total number of entries in $M$, denoted by $m$: 
$n \ll m$. 
The entries in $M$ are non-negative integers representing counts
(although our methods can apply to data with arbitrary values). 
We generate a differentially private table $M'$ from $M$ 
via the geometric noise 
mechanism \cite{Ghosh:Roughgarden:Sundararajan:09}.
This adds independent random noise to each entry in $M$, with the
distribution 

\centerline{$\displaystyle
 \Pr[ X = x] = \frac{1-\alpha}{1+\alpha} \alpha^{|x|}, $}
where $x \in \mathbb{Z}$, $\alpha = e^{-\epsilon/\Delta}$, $\epsilon$ is
the differential privacy parameter and $\Delta$ is the sensitivity 
parameter, for disjoint counts $\Delta=1$.

Because the noise can be negative, 
$M'$ can contain negative entries.
In many applications, it is not meaningful to have negative counts.
However, we can adjust for this, e.g., by rounding
negative values up to the smallest meaningful value, 0. 
Let $M'_+$ denote the anonymized table obtained from $M'$ via this procedure. 
For small range queries or for point queries, using $M'_+$ tends to be more
accurate than $M'$.
However, since the noise is symmetric, retaining negative values is
useful for large queries which touch many entries, since the noise
cancels. 
More precisely,
the sum of noise values is zero in expectation (but has non-trivial variance).

\begin{figure}[t]
\centering
\includegraphics[width=0.5\columnwidth]{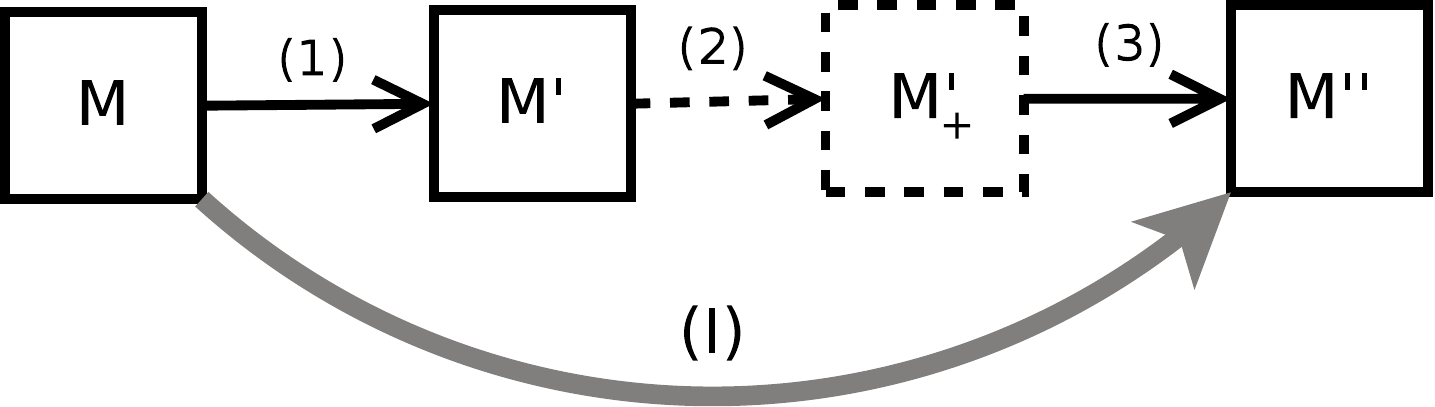}
\caption{Anonymization process: (1) adding Geometric/Laplace noise,
  (2) filtering negative entries, (3) sampling, (I) shortcut to
  generate $M''$ directly from $M$.}
\label{fig:sampling-model}
\end{figure}

Our aim is to publish $M''$, a compact summary of $M'$ (or of $M'_+$). 
The naive approach is as follows: 
 Compute the entire $M'$, by generating and adding noise 
for every entry in the (large, sparse) $M$, then summarize $M'$ to
obtain $M''.$  
This is costly to the point of impracticality, even though
$M''$ is expected to be relatively small, i.e., $\Theta(n).$
Instead, we compute $M''$ directly from $M$, without
materializing the intermediate table $M'$.
We illustrate this process schematically in Figure
\ref{fig:sampling-model}, which contrasts the {\em laborious} approach
to our proposed {\em shortcut} approach. 
The following observation is crucial for designing our algorithms.

\begin{remark}\label{binomial} 
For all summarization techniques we consider, each zero entry in $M$
has the same probability  $p$ to be chosen in $M''$. 
This is since the summarization methods depend only on the values of
the items in $M'$, not their positions; and since 
the noise added to each zero entry is drawn from the same 
distribution.   
Hence, out of the $m-n$ zero entries of $M$, 
the number chosen for the summary $M''$
follows the Binomial distribution $\bin(m-n, p).$
\end{remark}

In the subsequent sections, 
we develop shortcut approaches for several summarization methods.
Each method requires careful analysis, to show that the resulting
$M''$ is distributed as if it were generated via summarization of
$M'$, and that the shortcut approach can be implemented efficiently. 
We also state how to use the resulting summaries to accurately
estimate common queries (range and subset queries), and how to choose
parameters such as sample size.

To simplify notation, we treat 
 $M$ as a (long) one-dimensional array.
However, the techniques directly apply to other data layouts, 
  since they consider each entry in the input independently. 
We interchangeably refer
to entries in $M$ as items, consistent with sampling
terminology. 
Thus, we say that $M$ has $m$ items, of which $n$ have non-zero count.  
For clarity of exposition, 
we assume that the values $n$ and $\|M\|_1 = \sum_{i} |M(i)|$ are not 
sensitive. Otherwise, it is straightforward to
add the necessary noise to mask their true values.

\subsection{High-pass Filter}
The simplest form of data reduction we consider is to apply a
high-pass filter to $M'$: 
this has the effect of retaining the large values, but replacing the
small values with 0. 
If we choose an appropriate cut-off to distinguish ``small'' from
``large'', the result is to zero out a large portion of the data,
meaning that the remainder can be described compactly. 

Let $\theta$ be the cut-off value
($\theta \geq 0$). If an item $M'(i) \geq \theta$, then $M''(i) = M'(i)$, 
else we set $M''(i) = 0$. Our approach for generating $M''$ is to
consider the non-zero entries of $M$ separately from the zero-entries. 
First, in $O(n)$ time, we perform filtering for the non-zero entries in $M$:
generate and add noise, then determine whether to retain them. 
For the $m-n$ zero entries of $M$, this procedure is too slow (recall that $n\ll m$). Therefore, we
design a statistical process which achieves the same overall distribution over outputs, without explicitly
adding noise to each zero entry.

We now describe
our statistical process to filter originally zero entries. 
Let $p_\theta$ denote the probability that a zero entry
in $M$ passes the filter $\theta.$ 
We draw the number of entries $k$ that pass
the filter from the distribution $\bin(m-n, p_\theta)$, per
Remark~\ref{binomial}. 
We then choose uniformly at random $k$ zero entries in $M$. For
each such entry $i$, we generate the value $M''(i)$ by adding noise, {\em conditional} on the fact that the noise
 exceeds $\theta$. 
This may seem like a lot of effort to go to in order to generate noise
in the output data, but it is a necessary step: we must simulate exactly the output of filtering over the full 
table $M'$, in order to preserve the differential privacy guarantee. 
Algorithm {\sc Filter} summarizes the algorithm for this shortcut
process. 
We prove its correctness in Appendix~\ref{app:proofs}.

\para{Algorithm {\sc Filter}$(M)$:} 
Generates $M''$ via high-pass filter.

\newcounter{tc}\begin{list}{\arabic{tc}.}{\leftmargin=2em}\usecounter{tc}
\item 
For every non-zero entry $M(i)$, add geometric noise with parameter $\alpha$
to get $M'(i)$. 
If $M'(i) \geq \theta$, add $M'(i)$ to $M''$.

\item 
For zero entries, sample a value $k$
from the binomial distribution $\bin(m-n, p_\theta)$, where  
$ p_\theta \triangleq  \frac{\alpha^\theta}{1+\alpha} $.

\item
Uniformly at random select $k$ locations $i$ from $M$ such that
$M(i)=0$.
For each of these $k$ locations, 
include $i$ with value $v$ in $M''$ where $v$ is sampled   
according to the distribution
\[ \Pr[ X \leq x ] = (1 - \alpha^{x - \theta+1}) \]

\end{list}

\begin{theorem}
\label{thm:highpass}
Algorithm {\sc Filter} generates a summary with the same distribution as
the laborious approach under high-pass filtering with parameter
$\theta \geq 1$. 
\end{theorem}

\subsection{Threshold and Priority Sampling}
\label{sec:threshold_sampling}
More sophisticated data reduction techniques are based on sampling. 
In this section and the next, we discuss how to generate
$M''$ as a random sample of $M'$, without explicitly generating
$M'$. 
There are many sampling procedures.
The simplest is to uniformly sample items in $M$, then
create $M''$ by adding noise to them.  
This is easy to implement, but unlikely to have much utility: since
$M$ is sparse, we would sample almost exclusively 
 entries with $M(i)=0$, making the sample virtually useless. 
Instead, we extend the intuition from the high-pass filter: items in
$M'$ with high (noisy) values are more likely  
to correspond to non-zero entries in $M$, i.e., to original data. 
We should include them in our sample with higher probability. 
The filtering approach achieved this deterministically: items
below the threshold had zero chance of being included. 
When sampling, we now allow every item a chance of being in the summary, but
set the probability proportional to its (noisy) count. 

\para{Threshold Sampling.}
We first consider \textit{threshold
  sampling}~\cite{Duffield:Lund:Thorup:03}, and define the weight
$w_i$ of  
 item $i$ to be the absolute value of its noisy count $|M'(i)|.$ 
The method samples each entry with probability proportional to its
weight. Hence, an item with weight
$M'(i)$ has the same probability of inclusion as an item with weight
$-M'(i)$. 
The threshold sampling procedure with parameter $\tau$ is as follows: 
we include item $i$ in the sample with probability 
$p_i = \min(\frac{|M'(i)|}{\tau},1)$. 
This means that truly heavy items that have $|M'(i)| > \tau$ are included
in the sample with probability 1.

\para{Algorithm {\sc Threshold}.} 
Generate $M''$ via threshold sampling.
\newcounter{ttc}\begin{list}{\arabic{ttc}.}{\leftmargin=2em}\usecounter{ttc}
\item For every non-zero entry in $M(i)$, add  geometric noise to
  get $M'(i)$ and add it to $M''$ with probability $p_{i} = \min(\frac{|M'(i)|}{\tau},1)$. 
\item For the zero entries, sample a number $k$
from the binomial distribution $\bin((m-n), p_\tau)$, 
 where  
\[ p_\tau \triangleq  \frac{2\alpha (1 - \alpha^\tau)}{\tau(1- \alpha^2)} \]

\item 
Uniformly at random select $k$ entries $i$ from $M$ such that $M(i)=0$. 
For each of these $k$ entries, draw the summary value $M''(i)$
from the distribution $\Pr[X \leq \nu]$ given by:
\begin{align*}
\begin{array}{ll}
  \tau \alpha^{-\nu}C_\tau(1-\alpha), & \mbox{if } \nu \leq -\tau  \\
  C_\tau (- \nu \alpha^{-\nu}  + (\nu +1) \alpha^{-\nu +1} - \alpha^{\tau+1}), & \mbox{if }  -\tau < \nu \leq 0 \\
   \frac{1}{2} +  \alpha C_\tau(1 - (\nu+1)\alpha^{\nu} + \nu \alpha^{\nu+1}), & \mbox{if } 0 < \nu \leq \tau \\
   \frac{1}{2} + \alpha C_\tau(1 - \alpha^\tau - \tau\alpha^\nu (1-\alpha)) , & \mbox{if } \nu  > \tau
\end{array} 
\end{align*}
\noindent
where $C_\tau= \frac{1 } {2\alpha ( 1 - \alpha^\tau) }$ is a constant
depending on $\tau, \alpha$.
\end{list}

The following result is proven in Appendix~\ref{app:proofs}.

\begin{theorem}
\label{thm:threshold}
Algorithm {\sc Threshold} generates generates a summary with the same distribution
as the laborious approach under threshold sampling with parameter
$\tau > 0$. 
\end{theorem}
\para{Priority Sampling.}
\label{sec:priority}
\textit{Priority sampling} has been advocated as a method to generate
a sample of fixed size, with strong accuracy properties
\cite{Duffield:Lund:Thorup:07}.   
The sampling scheme is defined as follows. 
Each entry is assigned a 
priority $P_i = \frac{|M'(i)|}{r_i}$, where
$r_i$ is a random value chosen uniformly from the range (0,1].
The scheme draws a sample of size $s$ by picking the items with the
$s$ largest priorities, and also retaining the $(s+1)$th largest
priority for estimation purposes. 

To efficiently build a priority sample of $M'$, suppose that we knew 
$\tau_s$, the $(s+1)$th largest priority. 
Then $M'(i)$ is sampled if
\[P_i = \frac{|M'(i)|}{r_i} > \tau_s \iff r_i < |M'(i)|/\tau_s \]

Since $r_i$ is uniform over (0,1], the probability of this event is
$\min(|M'(i)|/\tau_s, 1)$. 
In other words, this procedure can be seen as equivalent to
Threshold Sampling with threshold $\tau_s$. 

The crucial difference is that $\tau_s$ is data-dependent, so we do not know it in advance.
However, we can {\em guess} a good value: one which will yield a sample of size $s'=\Theta(s),$
where the constant hidden by the $\Theta$ notation is small (but $\geq 1$). 
We first use our guess for $\tau_s$ to draw a corresponding threshold
sample. We then augment each item in the resulting sample with an assigned
priority $r_i$.  
That is, conditional on item $i$ being in the sample, we draw an $r_i$
consistent with this outcome. 
For $i$, we must have $0 < r_i \leq |M'(i)|/\tau_s$ (else $i$ would not
have passed the threshold), but beyond this there are no additional
constraints, so $r_i$ is picked uniformly in the range $(0,
|M'(i)|/\tau_s]$, and  subsequently its value is held fixed. 

Now that we have generated the priorities for our sample of size $s'=\Theta(s)$ 
(which is distributed the same as the distribution of the $s'$ largest priorities over all of
$M'$), we can reduce the sample size to $s$ by picking the $s$ largest
priorities (and retaining the $(s+1)$th priority). The running time is
$O(s') = O(s)$ (in expectation).
The result has the desired distribution; see the proof in
Appendix~\ref{app:proofs}.  

\begin{theorem}
\label{thm:priority}
We can generate a priority sample of $M'$ via the shortcut approach,
in expected time $O(s+n)$.
\end{theorem}

\subsection{Combining Sampling with Filtering}\label{filter-sample}
We have argued that both sampling and filtering are useful ways to
summarize data. 
In particular, filtering removes small counts which are very likely noise.
Setting the filter threshold low can remove a lot of noise, but will
still pass too many items, while setting it too high will remove too
many true data items. 
A natural compromise is to combine sampling with filtering: consider
generating $M''$ from $M'$ by first filtering out low frequencies, and
then sampling the result.  
We expect this to give us the best properties of both
summaries: noise removal and bounded output size. 
In Appendix~\ref{app:samplefilter}, we analyze the combination of the 
high-pass filter with threshold sampling, and design a shortcut
algorithm which correctly computes this combination.

\section{Improving Accuracy on Range Queries}
\label{sec:dyadic}
Range queries that touch many entries tend to have much higher error
than small queries. 
Although in expectation the sum of noise values is 0 (so query answers
are expected to be correct),  
its variance is linear in the number of entries touched by the
query~\cite{Dwork:06}. 
In practice, the observed errors tend to be proportional to the
standard deviation (i.e., the square root of the number of cells touched). 

A natural way to make range queries more accurate is to publish anonymized
data at multiple levels of granularity, so that any range can be
decomposed into a small number of probes to the published data.
For one dimensional data, the canonical approach is {\em dyadic
  ranges}: Build a (binary) tree over the domain of the data. For each
leaf, count the number of data values 
in the interval corresponding to that leaf. For each internal node
$u$, compute the sum of counts over all the leaves in 
$u$'s subtree. 
We can then publish these counts, at all levels, in a privacy-preserving manner. Let $h$ denote the height
of this tree. In general, $h=\log m$.
Each individual's data now affects $h$ counts, i.e., all the node counts on the path from the individual's leaf to
the root. Hence, the sensitivity increases by a factor of $h$, and the noise in each count is higher. 
It is well known that any range query can be answered as a sum of at most $2h$ counts (at most two node counts per level). 
For large enough ranges, the higher noise added to each count is
countered by the smaller number of counts touched by it. 
Related ideas were developed
in~\cite{Xiao:Wang:Gehrke:10}: 
The authors compute the wavelet transform of the 
data and add noise to its coefficients. 
This approach also has a sensitivity of $\log m$, and answers range queries 
from $\log m$ counts, but with different constants. 
These schemes and their variants are described and analyzed in greater
detail in recent
work~\cite{Li:Hay:Rastogi:Miklau:McGregor:10,Hay:Rastogi:Miklau:Suciu:10}.  
In theory and in practice the results for dyadic ranges and wavelets
are quite similar: 
the variance of estimators for range queries under both techniques is
proportional to $O(\log^3 m)$.

However, a limitation of these approaches is that they make no attempt
to bound the amount of data.  
If anything, they may blow up the data size.
Thus, when $m \gg n$,  we face the same scalability problem.
Our solution is to combine these variants of private data publishing
with summarization: The counts computed by dyadic ranges or wavelets on the original data become the input data $M$. We 
then apply the shortcut approaches from the previous sections.

\begin{theorem}
\label{thm:dyadic}
We can generate summaries of size $s$ 
with the same distribution as the laborious
approach applied to dyadic ranges or Haar wavelets 
under high-pass filtering and threshold/priority sampling in
(expected) time $O(s + n \log m/n)$.
\end{theorem}

This approach extends naturally to multiple dimensions of data. 
However, care is needed: the sensitivity grows exponentially with the
number of dimensions, as $\log^d m$ (when we have $d$ dimensions
divided into $m$ ranges each). 
As each range query is answered by summing $O(\log^d m)$ counts, there are fewer
queries benefiting from dyadic ranges or wavelets, as their sizes must 
increase rapidly with the dimension. 

\para{Consistency Checks for Dyadic Ranges.} 
We can further refine results for range queries by using the 
inherent correlation between counts along  the same leaf to root path. 
For dyadic ranges over the original data, a node count 
is never smaller than the counts in the node's descendants. 
Therefore, it is natural to try to enforce a similar condition
in the summary data, by modifying some entries after publication.  

For filter summaries over dyadic ranges, we impose the following
post-processing step. 
If a node $u$ is selected in the summary $M'',$ but
at least one of $u$'s ancestors $v$ is not selected, then we drop $u$
from $M''$.
The intuition for this is straightforward: 
Since $M(v)\geq M(u)$, $v$ had a higher chance to pass the filter after
noise addition. 
The fact that it did not is strong evidence that its count (and thus
$u$'s count) is small, and likely zero. 
The evidence of $v$'s absence from the summary trumps the evidence of
$u$'s presence, so we drop $u$. 

It is not clear that such a condition is meaningful on the output of
sampling: 
the summary may omit some nodes $v$ as part of the random sampling, 
so the absence of some node does not give a strong reason for dropping
its descendants.
Consequently, we do not advocate imposing this condition on the output
of sampling.  
However it is meaningful to apply for the combination of 
filter and priority sampling described in Section~\ref{filter-sample},
if we do so after filtering but before sampling. 
This means that we have to work on the output of the filtering, which
may still be large, and so we do not obtain a result that directly
generates the final summary. 
However, this may be an acceptable quality/efficiency tradeoff for some data.

\section{Experimental Study}
\newlength{\figwidth}
\setlength{\figwidth}{0.33\textwidth}
\begin{figure*}[t]
\subfigure[{Filtering: Varying sample size}]
{\label{fig:exp1_filtering}\includegraphics[width=\figwidth]{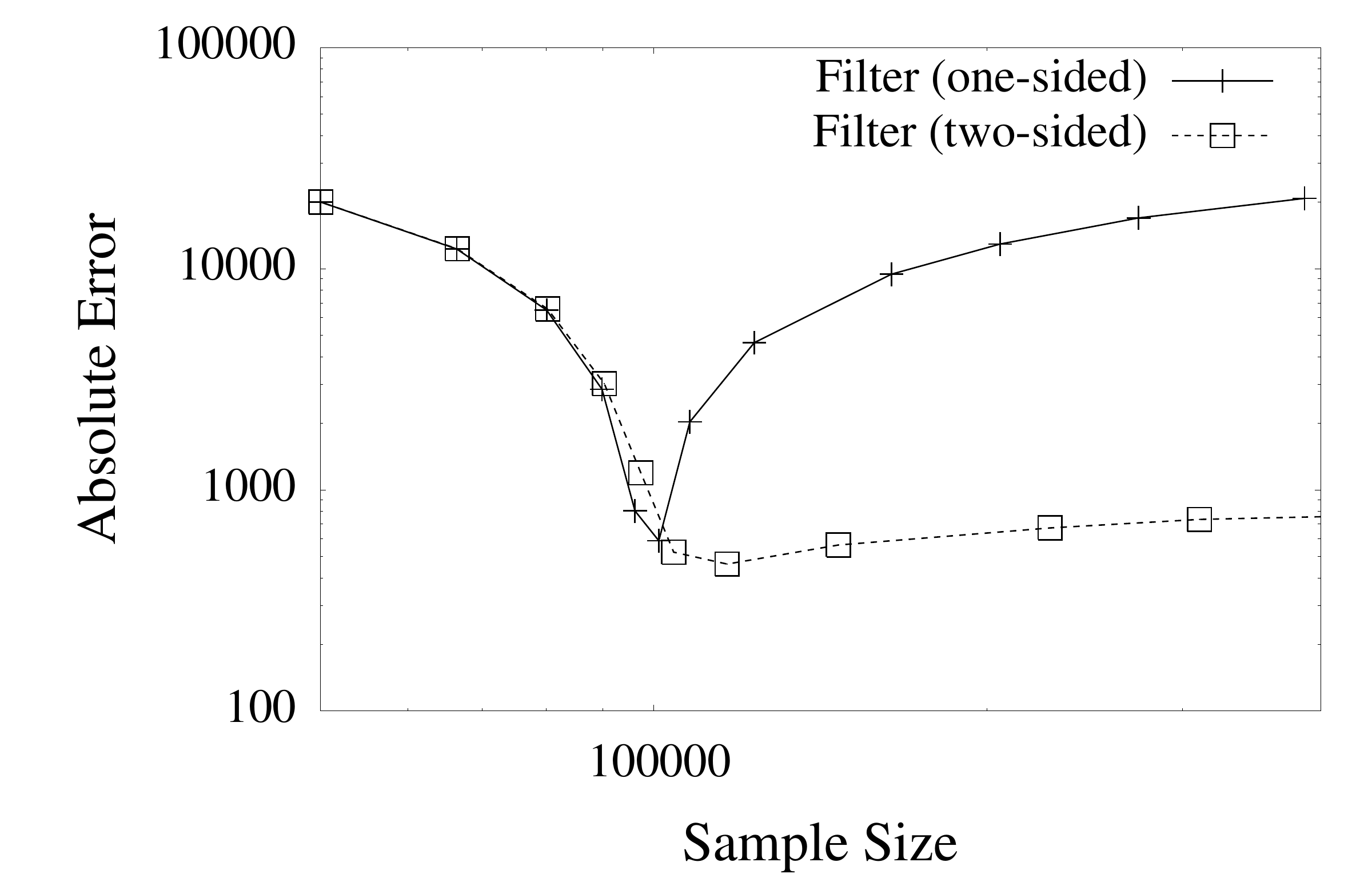}}
\subfigure[{Threshold and Priority sampling}]
{\label{fig:exp1_priority}\includegraphics[width=\figwidth]{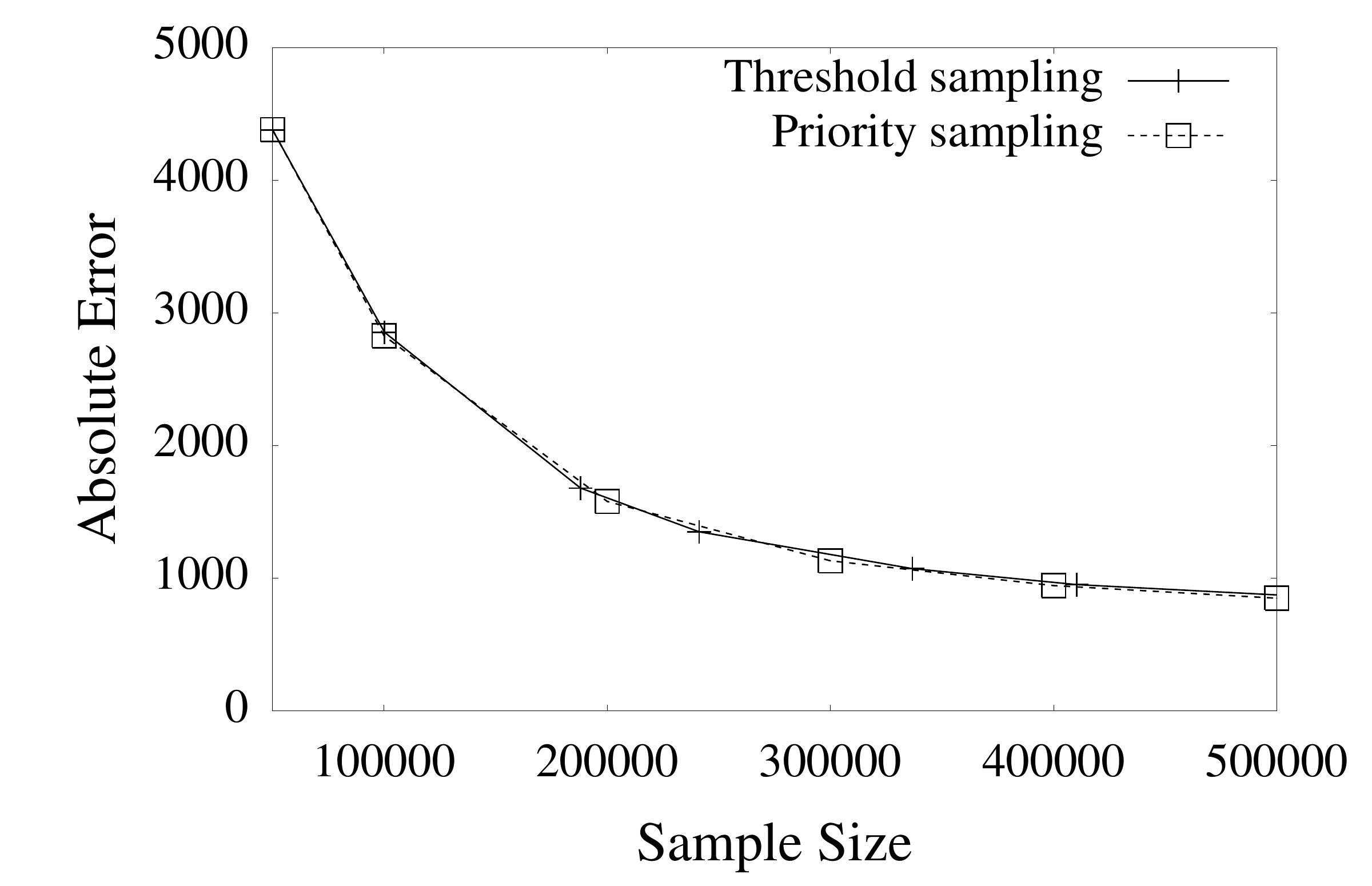}}
\subfigure[ {Filter-Priority sampling}]
{\label{fig:exp1_thresholdpriority}\includegraphics[width=\figwidth]{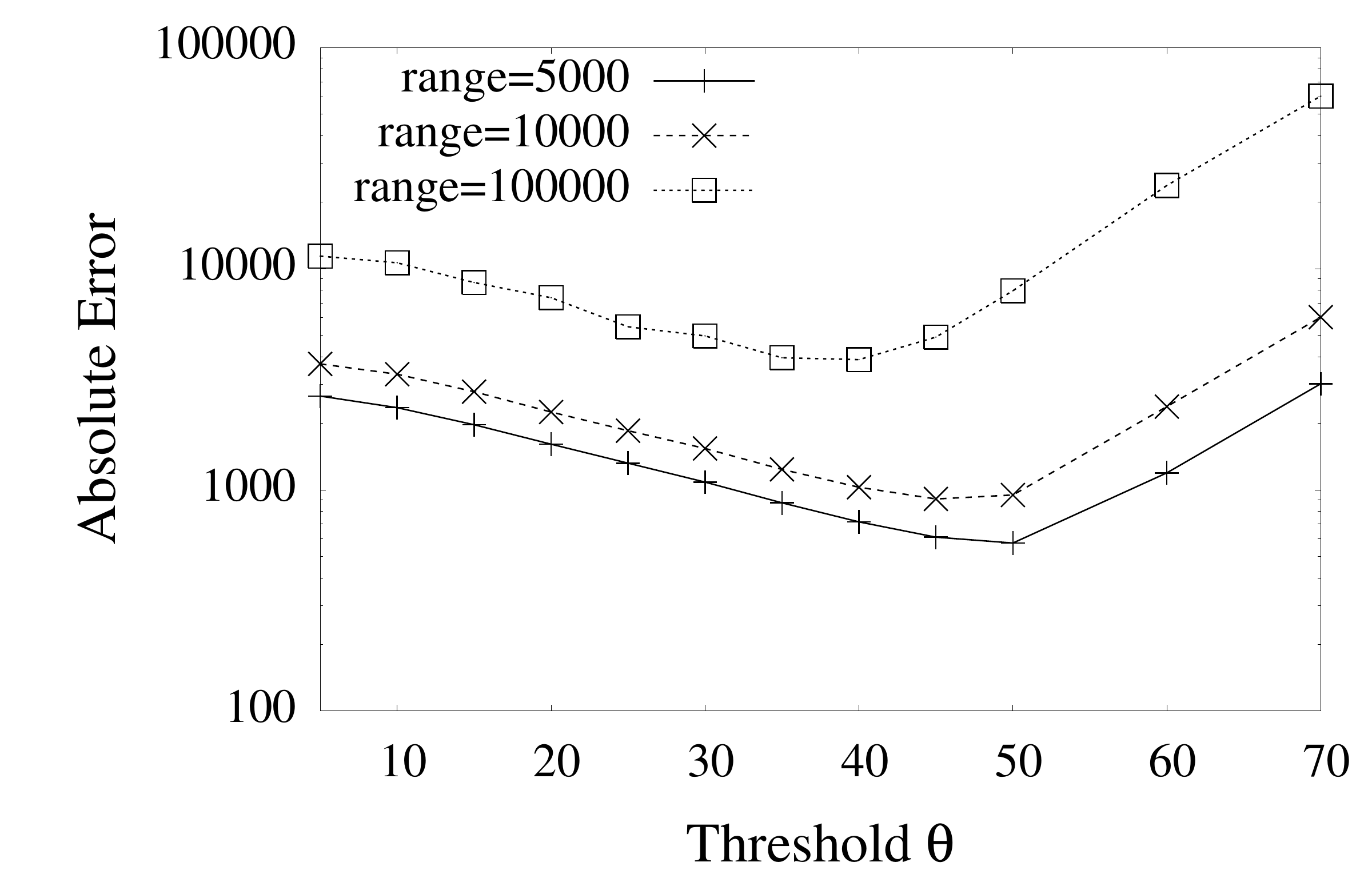}}
\caption{Impact of parameter choice for each method}
\label{fig:parameter}
\end{figure*}
\setlength{\figwidth}{0.38\textwidth}
\begin{figure*}[t]
\centering
\subfigure[ {Small subsets, ($\mu$=100, $\sigma$=20, $\rho$=0.1)}]
{\label{fig:exp2_compare1_small}\includegraphics[width=\figwidth]{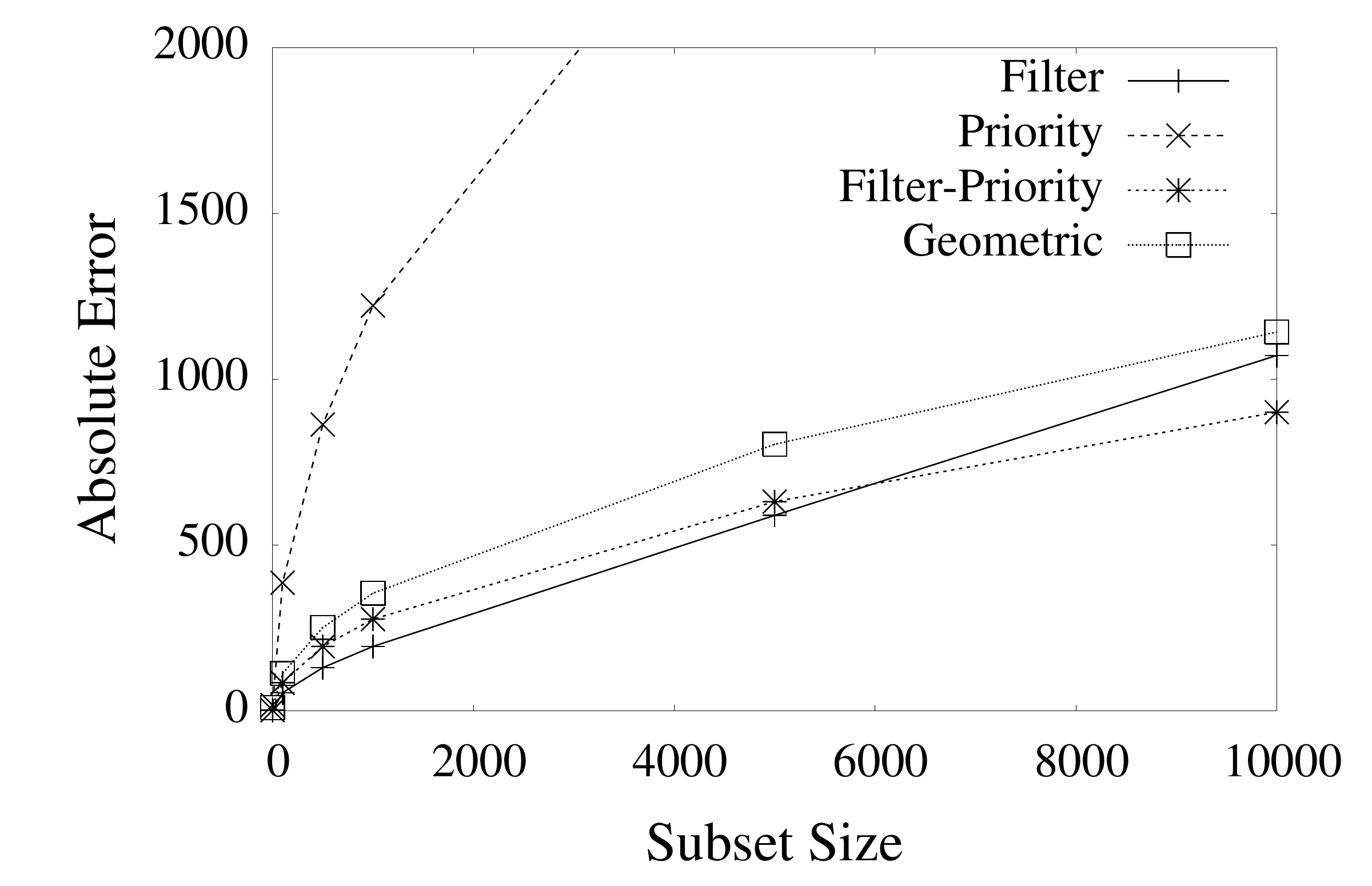}}%
\subfigure[  {Large subsets, ($\mu$=100, $\sigma$=20, $\rho$=0.1)}]
{\label{fig:exp2_compare1_large}\includegraphics[width=\figwidth]{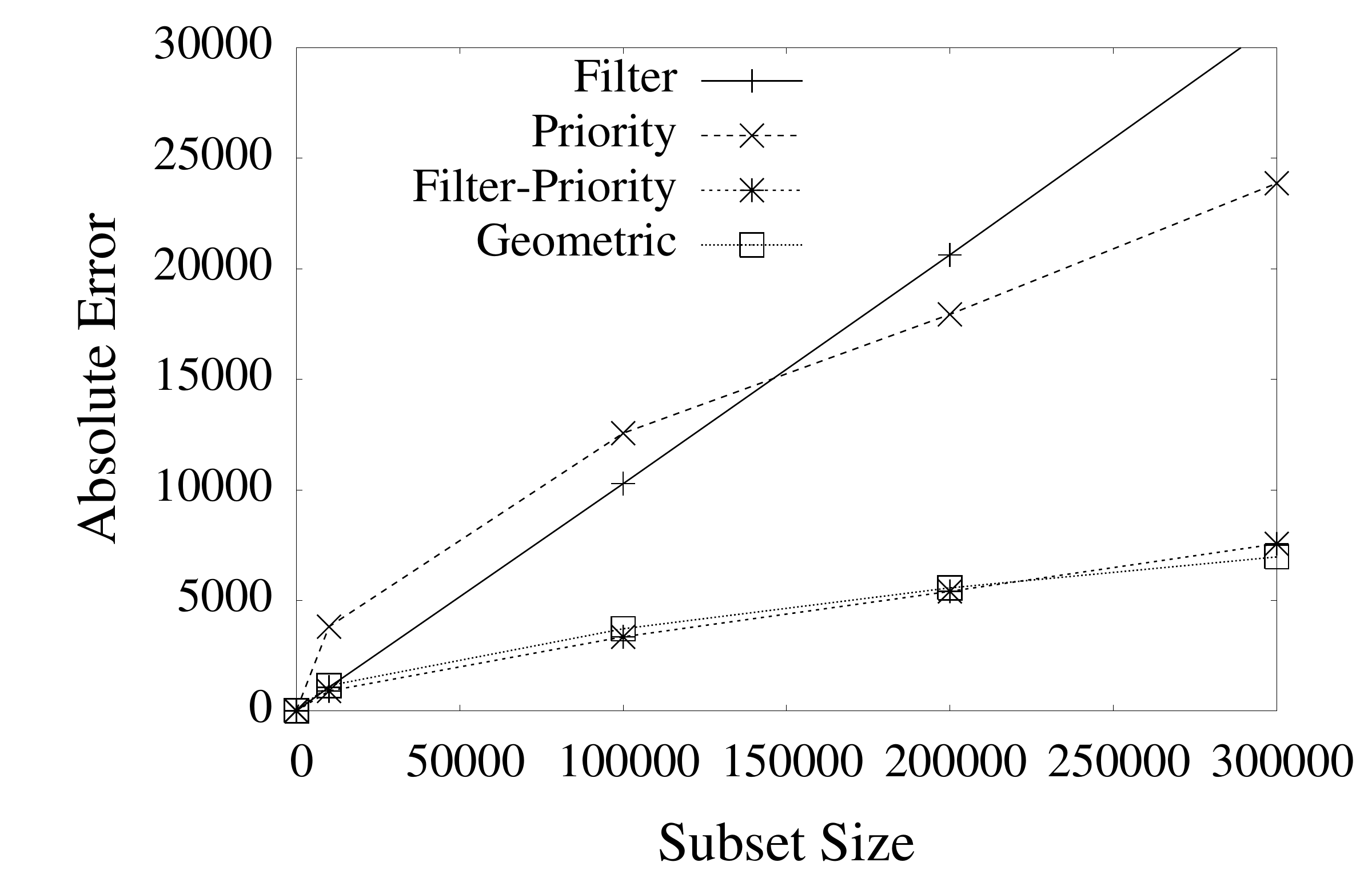}}

\subfigure[{Subset queries, ($\mu$=100, $\sigma$=40, $\rho$=0.1)}]
{\label{fig:exp2_compare2}\includegraphics[width=\figwidth]{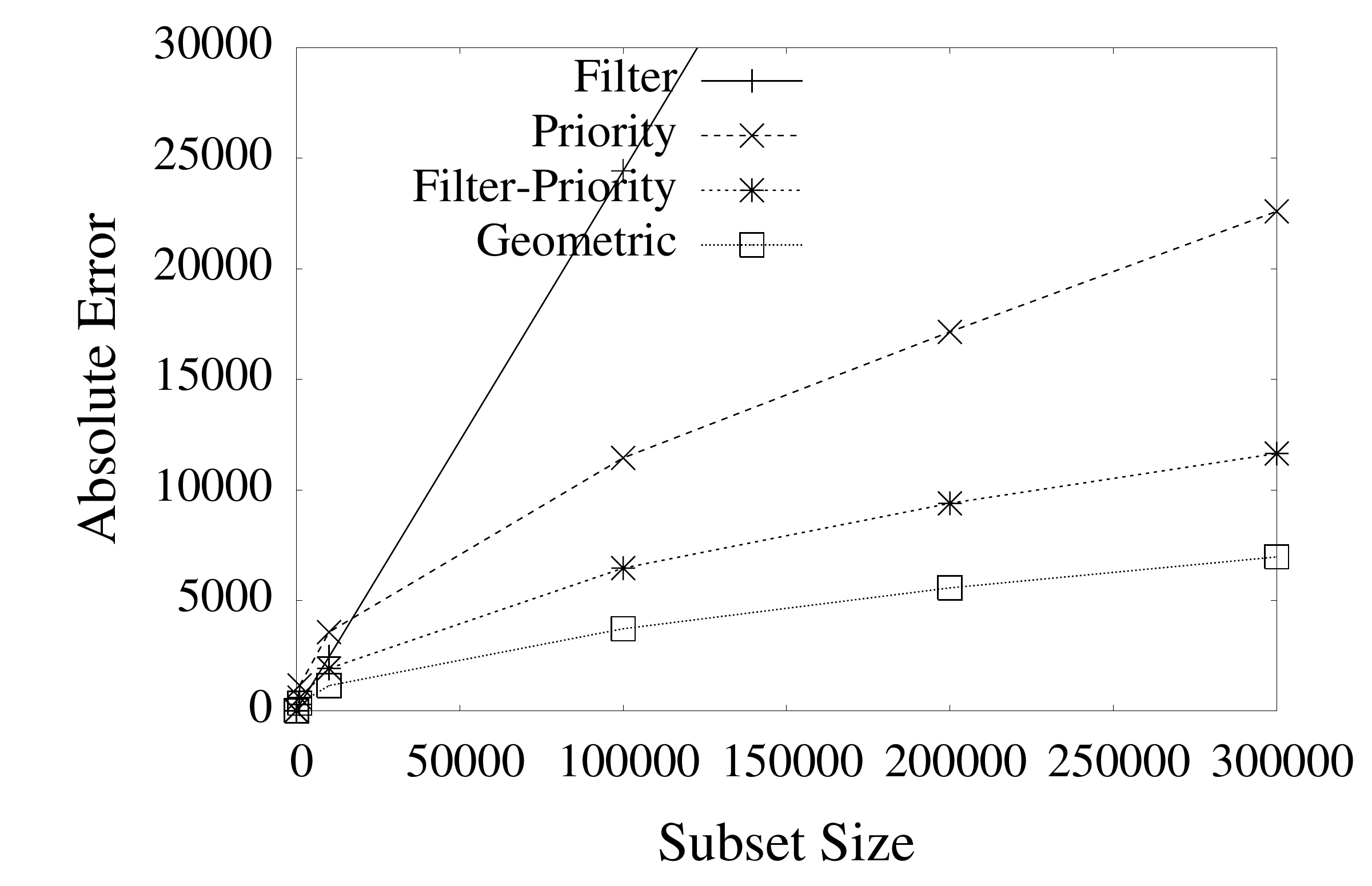}}
\subfigure[ {Subset queries, ($\mu$=100, $\sigma$=20, $\rho$=0.01)}]
{\label{fig:exp2_compare3}\includegraphics[width=\figwidth]{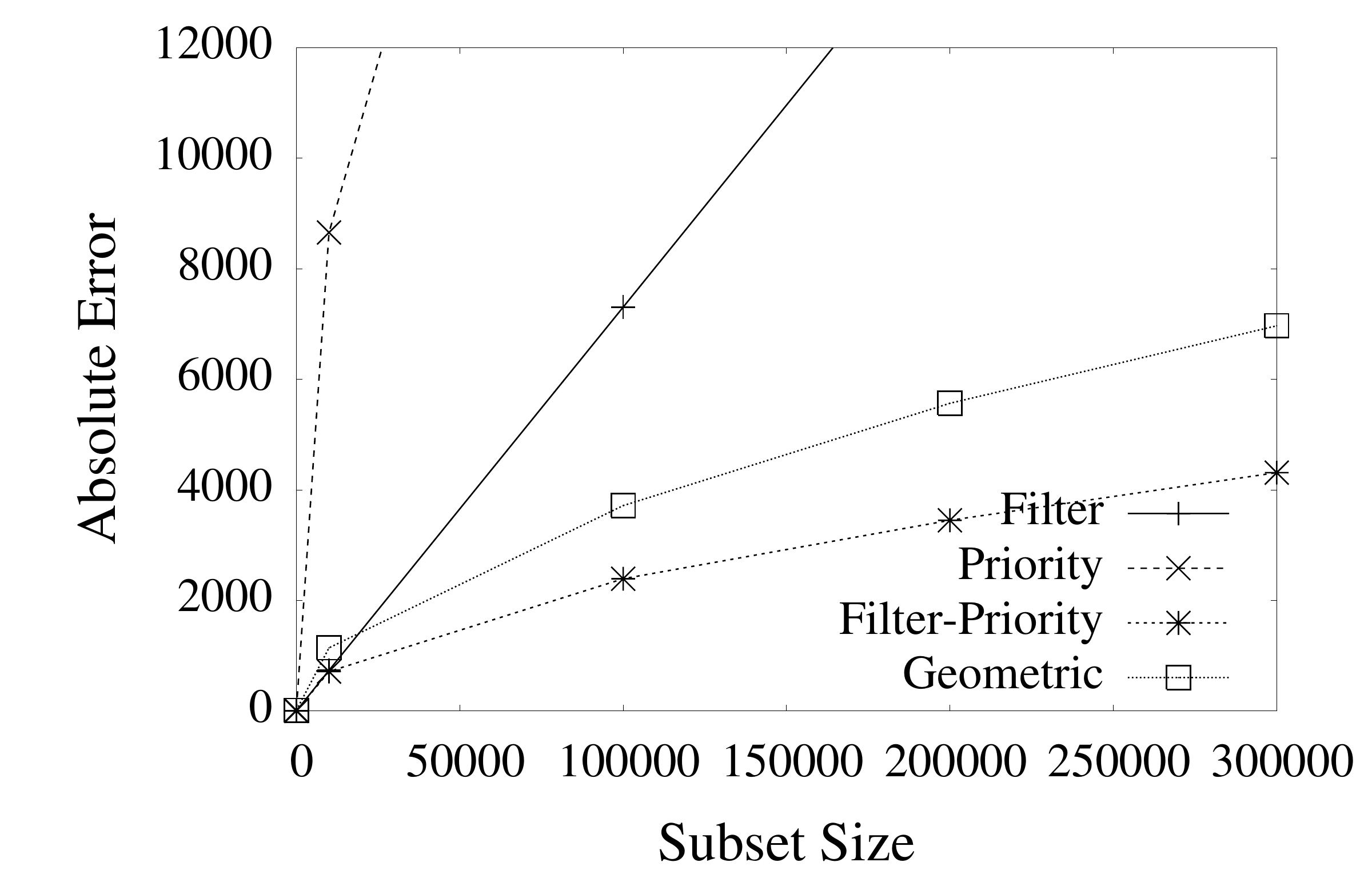}}
\caption{Experimental results on subset queries}
\label{fig:subset}
\end{figure*}

\subsection{Experimental Setup}
We first evaluate on synthetic data,
where we can observe the impact of varying data characteristics on
the quality of summaries produced.
We control parameters
including the  mean $\mu$ and standard deviation
$\sigma$ of the (non-zero) data values, 
the data density ($\rho = n/m$), 
and 
the distribution of locations of non-zero data values 
 in the input space, e.g., either uniformly scattered or skewed and
 concentrated in particular regions. 
Our experiments use an input size of $m=10^6$ as a default, so it
is feasible to compare to the solution of adding noise at each location.
The non-zero data entries are drawn from a Gaussian distribution
with mean $\mu=100$,  standard deviation $\sigma=20$, 
and they are then rounded to the nearest positive integer. 
The default data density is $\rho=0.1$, a moderately dense data set by
our standards, and the data values are uniformly distributed over the
data domain. 
The default privacy requirement is set to $\epsilon=0.1$.
We use
the standard geometric mechanism as the baseline to which we compare
our techniques, noting that this is not a practical solution for the
sparse data settings we consider. 
We measure utility by computing the absolute
errors for various queries, i.e. the $L_1$ difference between the
true query answer (on $M$) and that estimated from the published data ($M''$). 
This dimensionless quantity varies depending on the size of
the range and the nature of the data, but allows us to compare the
relative performance. 
We also report median relative errors over the query sets.
The queries are to compute the sum of a set of values  
specified as either a contiguous range of values in
the data space (range query), or as a subset of locations in the
domain (subset sum query).

\subsection{Filtering and Sampling Parameter Setting}
Figure \ref{fig:parameter} shows  
absolute error as
we vary data parameters ($\theta$ and $\tau$).
For these summaries, 
range queries and subset queries are 
effectively identical, so we do not distinguish them. 

\para{Filtering.} 
We compare the performance of the two high-pass filters:
one-sided (pass only positive values above $\theta$) 
and two-sided (pass all items whose absolute value is above $\theta$).  
We fix the range size $R=5000$ and
vary the filtering threshold $\theta$ to get different summary sizes.
Figure \ref{fig:exp1_filtering} shows the accuracy 
on a log scale, 
indicating that the accuracy varies by several orders of magnitude.  
Two-sided filtering is generally more accurate than
one-sided filtering. 
When most of the (originally) non-zero data points are in
the sample, query answers using two-sided filtering estimates are
unbiased. 
In this case
the relative error of this technique is very low: consistently around 1\%. 
One-sided filtering is more likely to  
over-estimate because some data points which were  originally zero
 are now positive entries in the sample. 
When the summary size is small, we expect that the summary is
dominated by the non-zeros and hence both techniques have similar
accuracy.
The minimal error occurs when the sample contains
most of the true data points and few originally zero entries.
But this may require a $\theta$ value that results in a larger than
desired summary. 
As we saw the same behavior across a variety of range sizes and
data types, 
we henceforth only use two-sided filtering,
since it is uniformly more robust.

\begin{figure}[t]
\centering
\includegraphics[width=\figwidth]{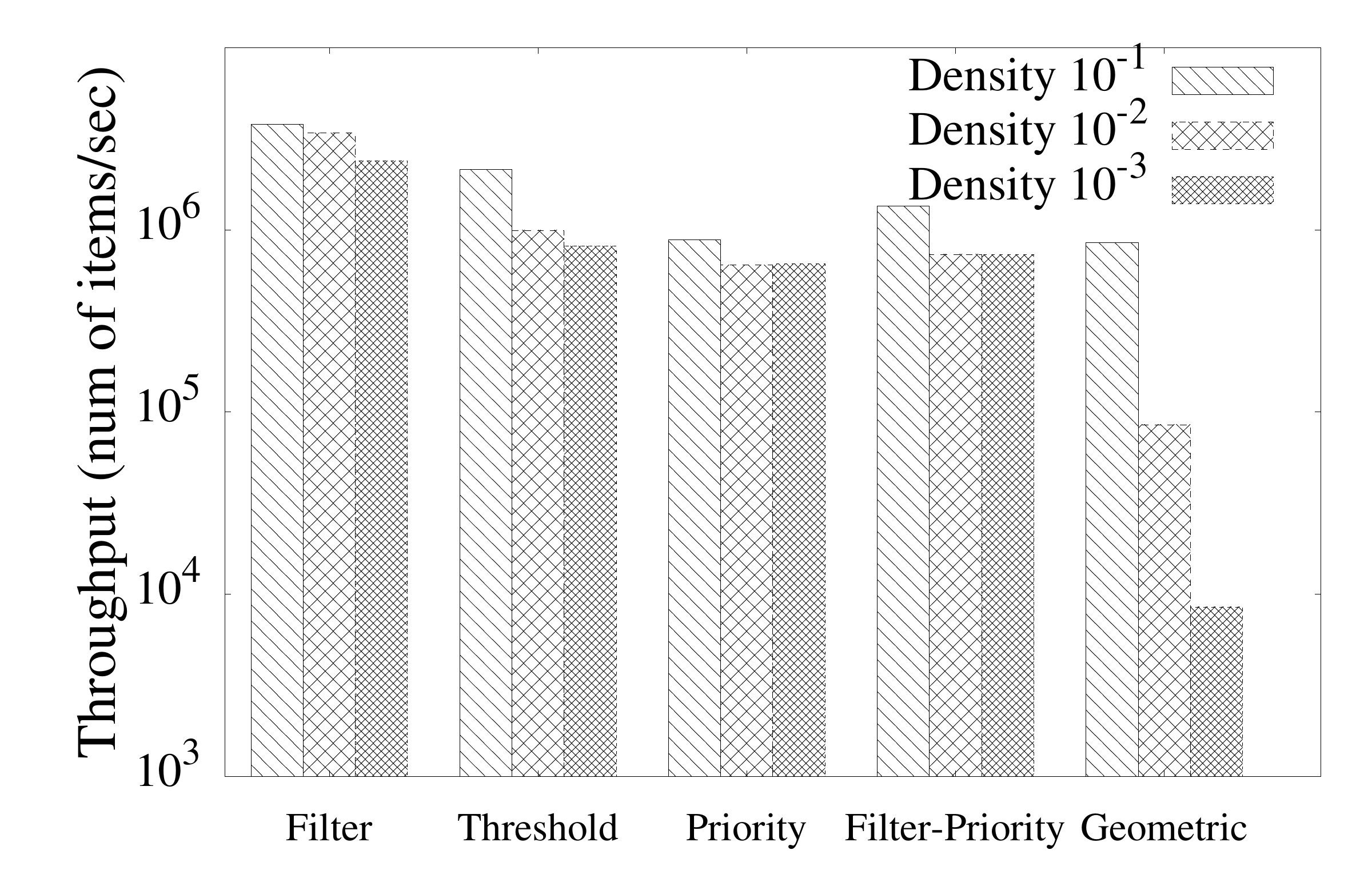}
\caption{Throughput as density varies}
\label{fig:exp2_throughput}
\end{figure}

\para{Threshold and priority sampling.} 
Figure~\ref{fig:exp1_priority} shows the
accuracy of threshold sampling and priority sampling
when we fix the range size $R=5000$ and set the parameter $\tau$ for
threshold sampling to get a desired sample size $s$ 
(see Section~\ref{app:threshold}).
This corresponds to relative errors in the range 1\% to 7\%. 
For priority sampling, the sample size $s$ is the input parameter.  
The behavior of
threshold sampling is very similar to that of priority sampling, as
expected from our analysis. 
The only difference is that priority sampling returns a sample with
fixed size $s$ as required, but needs more computation to find the
$(s+1)$th priority and adjust the weights accordingly, as is observed
 in Figure~\ref{fig:exp2_throughput}.  
Since threshold and priority sampling behaved similarly
 across all other range sizes tested, we show only the latter from now
 on. 

\para{Filter-priority sampling}. 
Figure~\ref{fig:exp1_thresholdpriority} shows the accuracy for the
combination of two-sided filtering with priority sampling
(which we dub ``filter-priority sampling'') for
 $\theta$ between 5 and 70.
We  fix the priority sample size to $s=10^{5}$, 
which is equal to the cardinality of the original input data. 
The relative errors measured are within [0.3\%, 6\%]. 
For this setting, we find that $\theta=40$ is about optimal. 
This is in the region of half of the mean $\mu$ of the non-zero data
values. 
In general, the threshold
$\theta$ should maximally filter most of the (upgraded) zeros but
still retain a large number of the original non-zero data, 
hence it should be neither too large nor small. 
For larger range queries, we want to set
$\theta$ smaller to make sure most of the non-zeros appear in the
sample, since on average a range query contains more non-zeros and
this technique relies on the priority sampling to give unbiased
estimations to query answers. 
The choice of $\theta$ is also affected by the data density. 

\setlength{\figwidth}{0.33\textwidth}
\begin{figure*}[t]
\subfigure[ {Dyadic ranges (uniform data)}]
{\label{fig:exp3_dyadic1}\includegraphics[width=\figwidth]{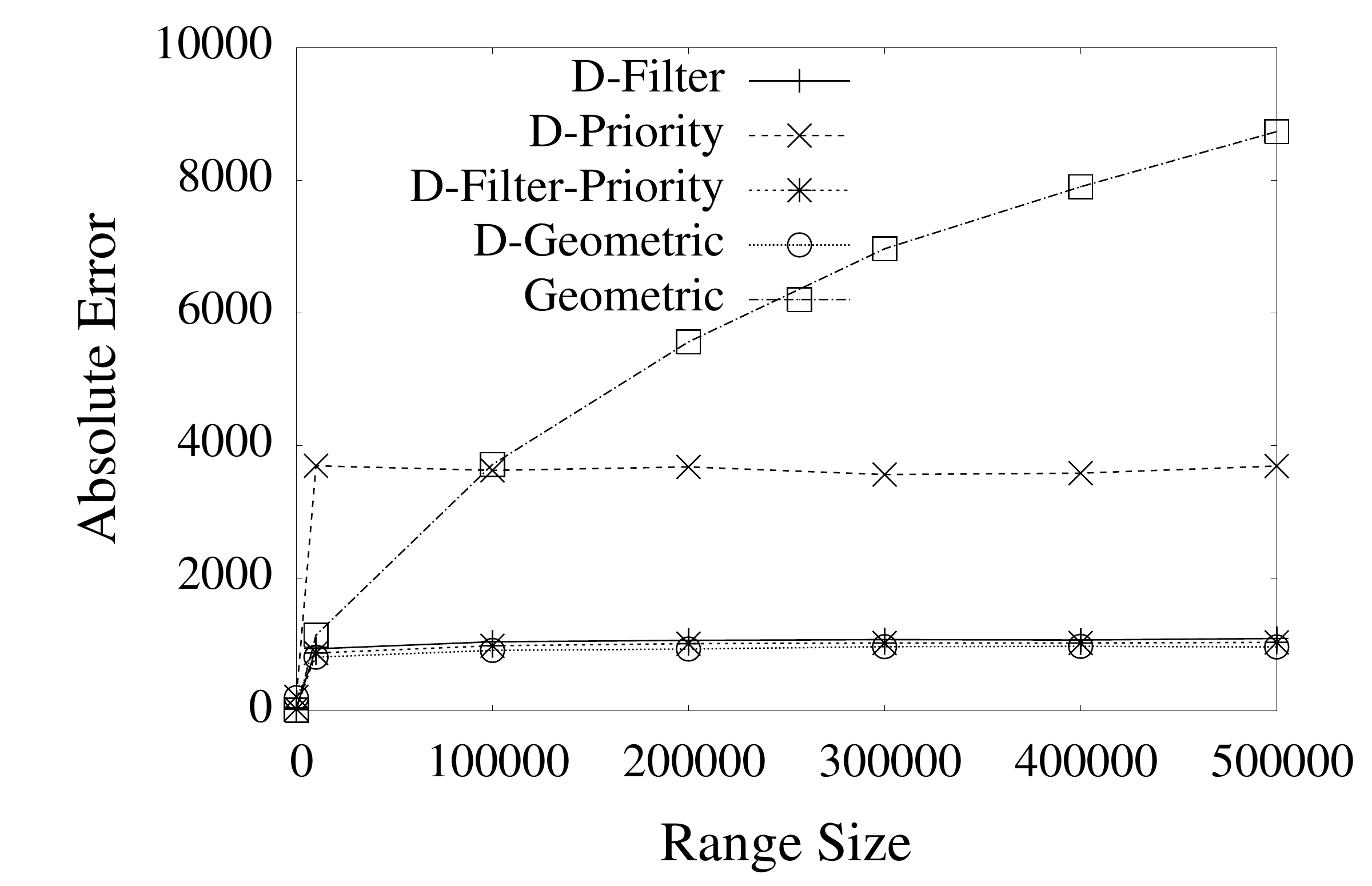}}
\subfigure[ {Dyadic ranges (skewed data)}]
{\label{fig:exp3_dyadic2}\includegraphics[width=\figwidth]{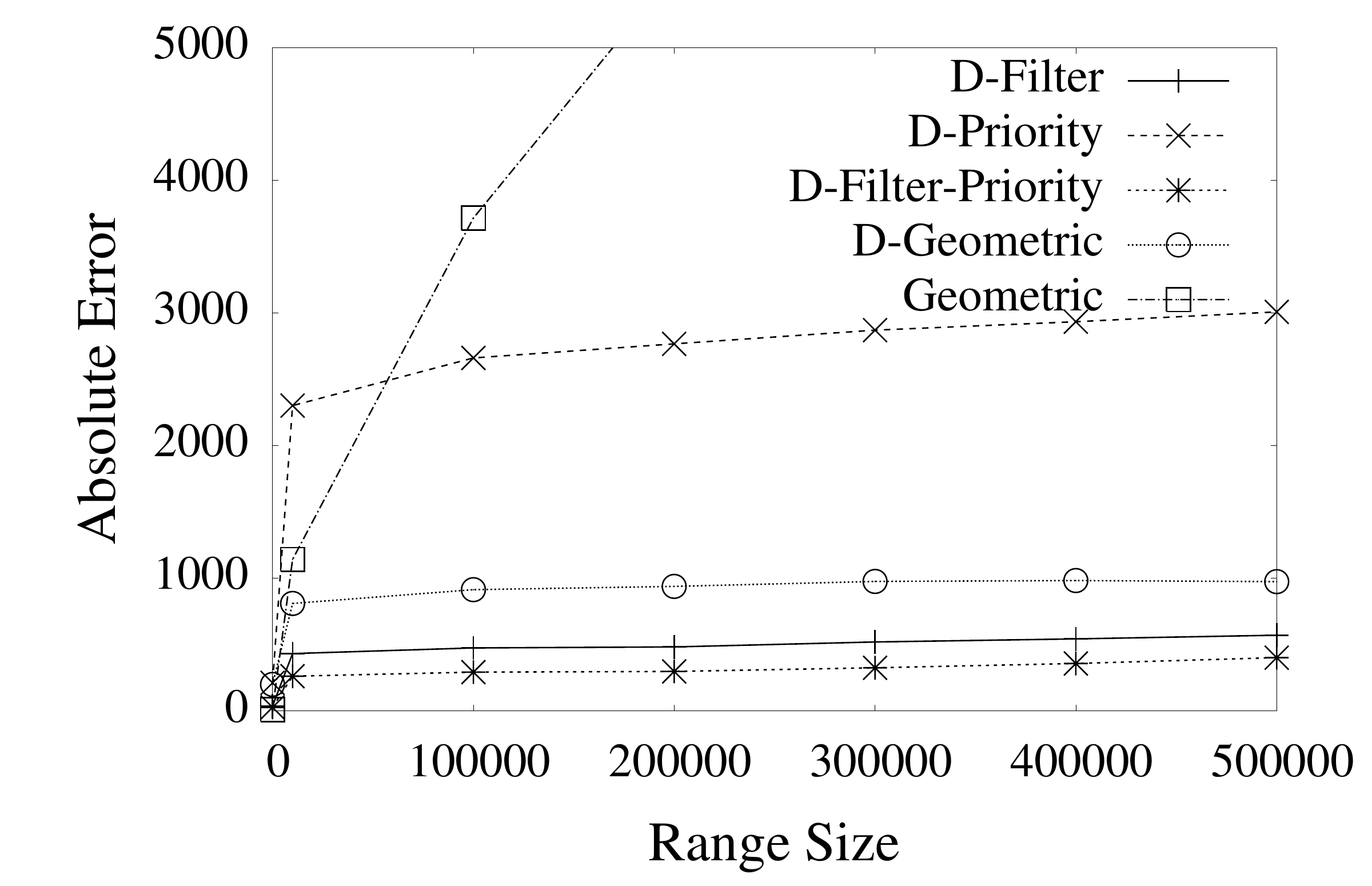}}
\subfigure[ {Dyadic ranges (skewed data, consistency)}]
{\label{fig:exp3_dyadic_consistency}\includegraphics[width=\figwidth]{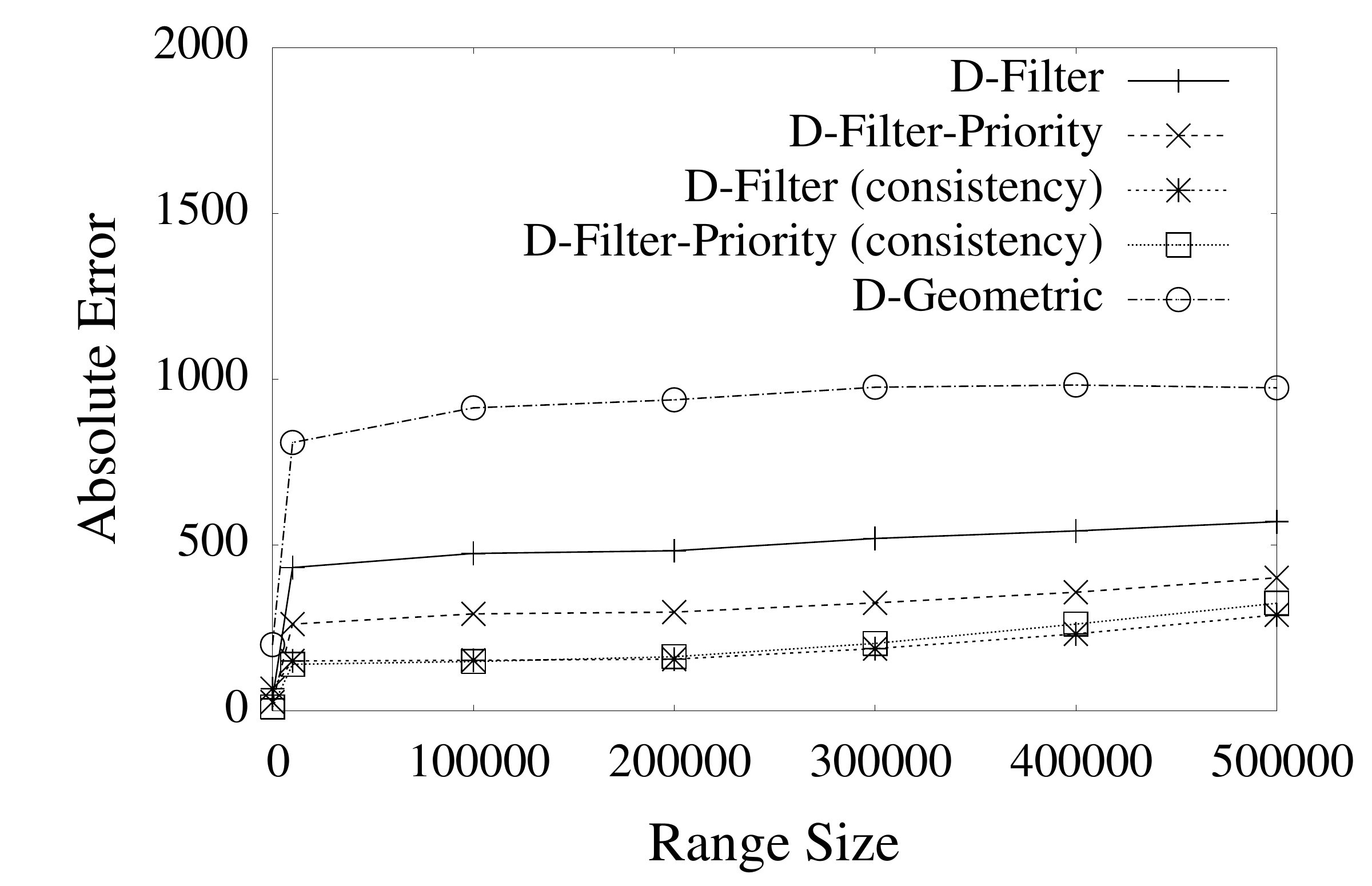}}
\caption{Impact of Dyadic Ranges for query answering}
\label{fig:dyadic}
\end{figure*}

\subsection{Comparing the Summarization Techniques}
We next create summaries of size $s=10^{5}$, and compare performance 
to the baseline geometric mechanism in Figure \ref{fig:subset}. 

\para{Range query accuracy.}
Figures \ref{fig:exp2_compare1_small} and
\ref{fig:exp2_compare1_large} show the accuracy of high-pass
filtering, priority sampling, filter-priority sampling, and the naive 
geometric mechanism, for both medium and large range queries using the default
dataset.   
Apart from priority sampling, all techniques have relative error
around 5\% for subsets of size 100, and this decreases as the subset
size increases. 

Note that the geometric mechanism publishes the entire noisy data
domain, which is typically infeasible, so 
we see how well we can do against this benchmark when
releasing a much smaller summary.  
For small and medium range
queries, filtering, with $\theta=50$, performs best while threshold
and priority sampling is preferable when the range size is large
enough. 
This also exhibits the same behavior as in
Figure \ref{fig:exp1_thresholdpriority}. 
Filter-priority sampling (here we use $\theta=40$) 
combines the advantage of the two techniques,
having consistent accuracy comparable to that of the
geometric mechanism.

This is a somewhat surprising outcome: at best, we hoped to equal the
accuracy of the (unfeasible) geometric mechanism with a compact summary; 
but here and elsewhere, we see examples with {\em better} accuracy
better than this benchmark.
The reason is that summarization is helping us: the noise
introduced by the privacy mechanism is less likely to reach
the output summary, so we get somewhat more accurate answers. 
This does not alter privacy guarantees---as the user
is free to do any post-processing of the output of the geometric
mechanism, such as apply a filter to it---rather, this indicates that
the summarization techniques maintain the same privacy and can help
improve utility of the released data.  

Figure \ref{fig:exp2_compare2} shows the corresponding plot with
higher standard deviation $\sigma=40$.
Filtering now performs much worse
as it is harder to separate the true data from the noise
introduced in $M'$. 
Threshold and priority sampling start to outperform filtering
over shorter ranges, around a thousand items in length. 
Filter-priority sampling, with a smaller threshold $\theta=20$,
is better than individual filtering or sampling techniques. 
In this experiment, the summary techniques are less accurate than the
geometric mechanism.  
The gap is narrowed for larger summaries: 
picking a sample size of $2\times 10^5$ (twice as large)
is sufficient to make the filter-priority summary equal to the 
geometric mechanism in accuracy. 

\setlength{\figwidth}{0.33\textwidth}
\begin{figure*}[t]
\subfigure[{Subset queries}]
{\label{fig:exp4_map1}\includegraphics[width=\figwidth]{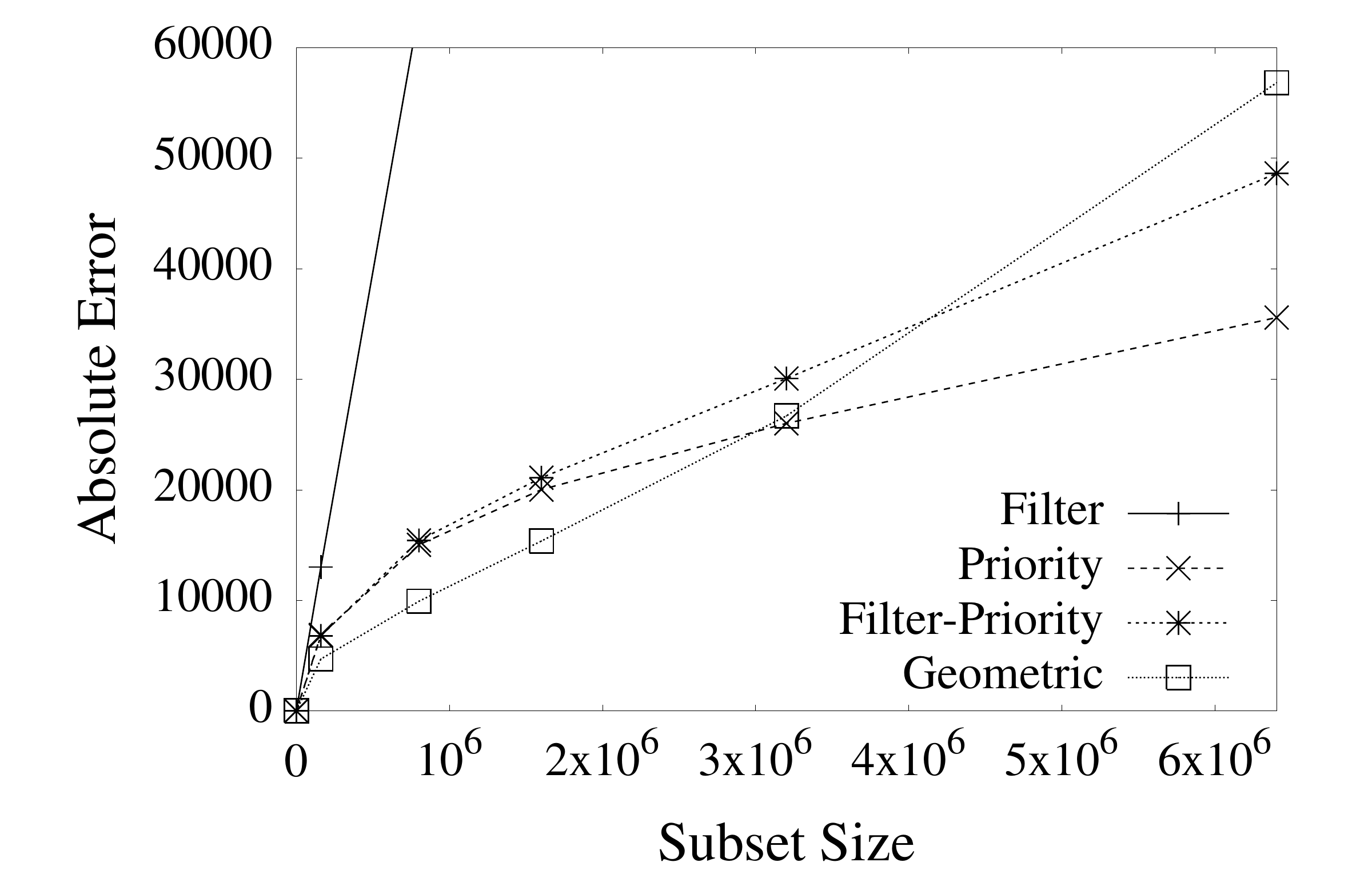}}
\subfigure[ {Dyadic ranges }]
{\label{fig:exp4_map2}\includegraphics[width=\figwidth]{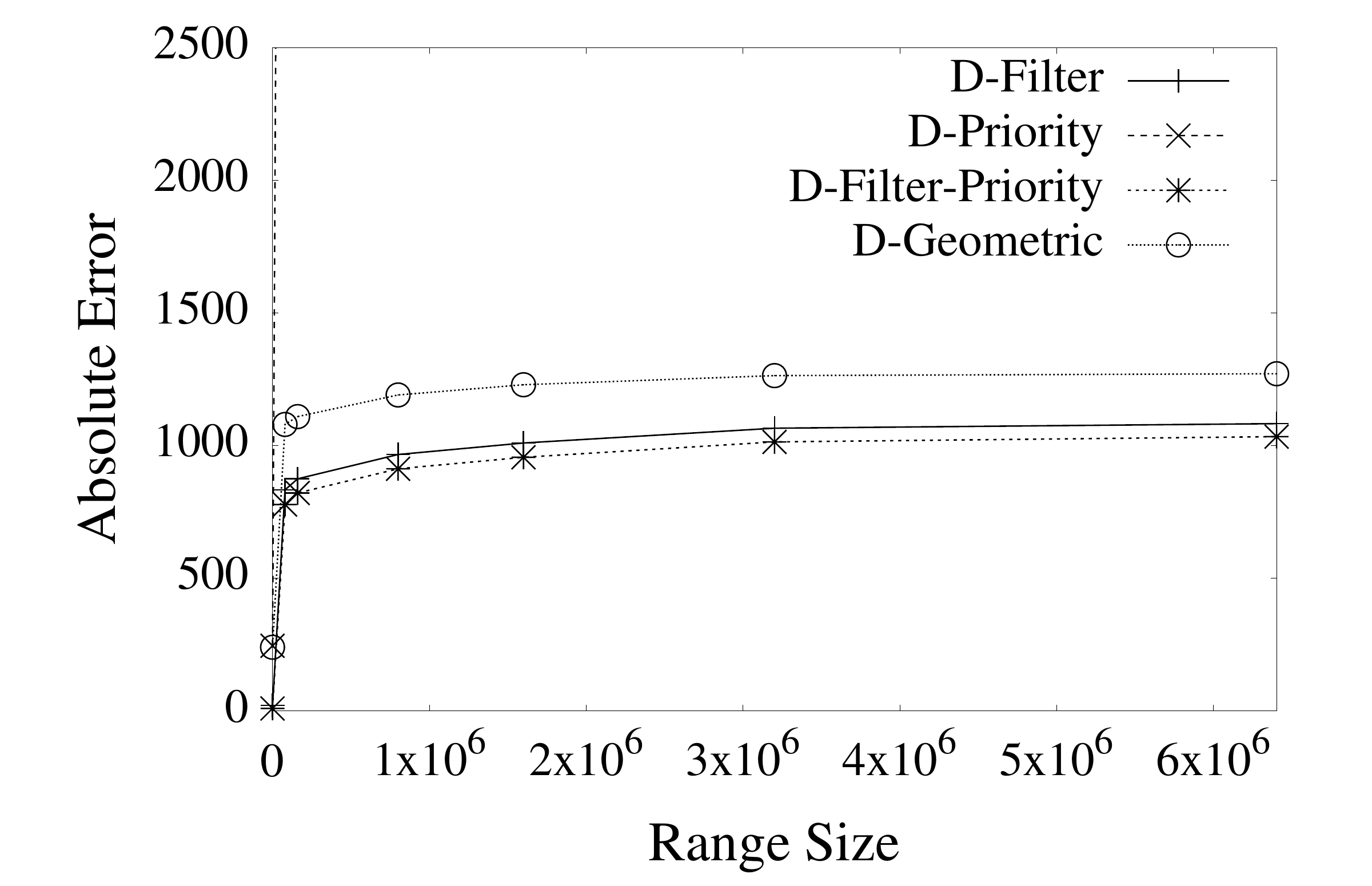}}
\subfigure[ {Dyadic ranges (with consistency)}]
{\label{fig:exp4_map_consistency}\includegraphics[width=\figwidth]{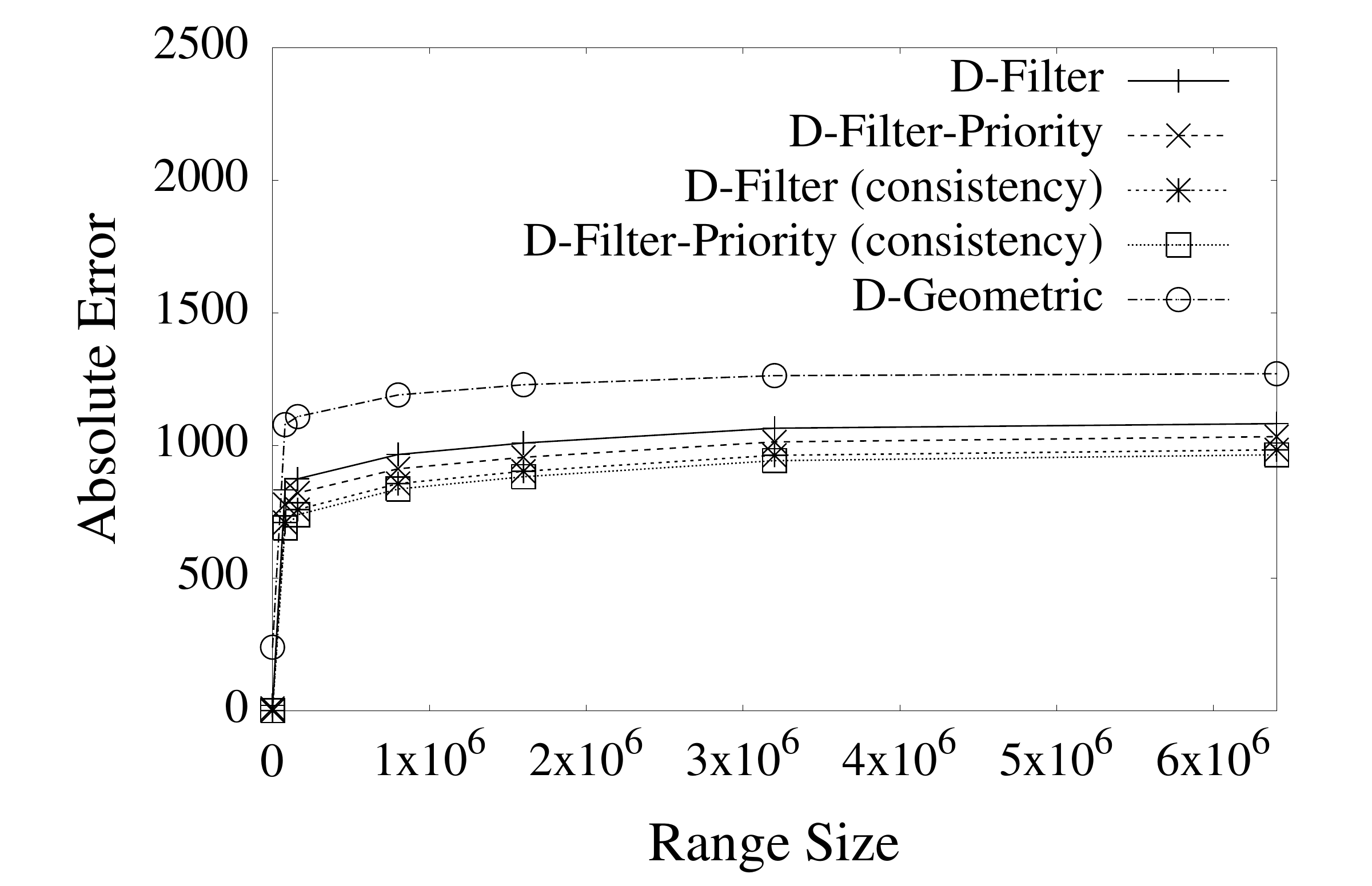}}

\caption{Experimental results using ``OnTheMap'' data}
\label{fig:onthemap}
\end{figure*}

\para{Impact of data density.}
Next, we reduce the data density to $\rho=0.01$ while using the same
input size $m$ and setting the sample size to be the number of non-zeros,
i.e., $s=10^4$. 
Figure~\ref{fig:exp2_compare3} shows 
that filtering outperforms sampling across a range of sample sizes, 
because with more zero entries, 
the total probability of a zero appearing in the
sample increases, and the proportion of true data points sampled
decreases. 
But filter-thresholding (with
$\theta=50$) has yet higher accuracy and outperforms the geometric
mechanism across the whole spectrum of subset sizes due to efficiently canceling much noise from zero entries. 
This suggests that, across a range of data types and query
sizes, this hybrid method has the best of both worlds: it uses
filtering to prune away pure noise, and  sampling to
favor more significant data entries without bias, producing a
summary of fixed size. 

\para{Throughput.}
Recall that our motivation is that data release via methods like the
geometric mechanism is not scalable when the data is
sparse: it produces a very large output (of size $m$), and may be very
slow to create the summary. 
We now justify our effort in creating compact summaries by observing
that they can be created very efficiently. 
We measure the throughput, or the number of (non-zero) entries processed per second,  of all techniques under different
data density values, $\rho=\{10^{-1}, 10^{-2}, 10^{-3}\}$. 
This spans the range of densities observed in Table~\ref{tblsparse}.
The size of input data domain, $m$, is held fixed,
and sample size $s$ is set equal to the cardinality of the original
data, $n$. 
Figure \ref{fig:exp2_throughput} shows 
that filtering and
threshold sampling are the fastest since they require less
computation.
Priority sampling has a higher cost due to the extra effort to find the $(s+1)$th priority and
adjusting the weights. 
Under the very densest setting, $\rho=0.1$, its
running time is about the same as the basic geometric mechanism. When
the data is more sparse, all of these techniques are faster than the
basic geometric mechanism, by orders of magnitude. 
This demonstrates that the laborious
approach to generating the samples, which requires materializing the
(assumed huge) $M'$, is infeasible in most cases. 

\subsection{Summarizing with Dyadic Ranges}
As noted in Section \ref{sec:dyadic}, answering range queries via
dyadic ranges has proven successful when it is feasible to publish the
full output of the laborious approach. 
We next combine the dyadic range technique 
with the filtering and sampling methods 
to produce a bounded output size, 
and evaluate the performance on range queries. 
Specifically, we consider filtering, priority
sampling, and filter-priority sampling on dyadic ranges, 
as well as the geometric mechanism both with and without dyadic
ranges. 
The size of the sampled dyadic ranges is set equal to number of
non-zeros in the input, $s=10^{5}$, and we vary the range sizes.

Figure \ref{fig:exp3_dyadic1} shows that
 priority sampling is
worst of the filtering and sampling techniques. 
This is since
the sensitivity of the dyadic ranges is large (i.e., $\log m$), hence
the noise added is large, giving zeros in dyadic ranges a high
probability to be in the sample. 
Filtering is preferable since it can filter most of original zeros;
 even if we drop some entries with very small magnitude, this
 does not dramatically affect the accuracy of the overall answer on
 the whole range. 
When range size 
$R \geq 10^4$ (corresponding to 1\% of the data space), 
using dyadic ranges has better accuracy than without them. 
After this point, the accuracy depends only very weakly on the range size,
since each query probes about the same number of entries 
(i.e., twice the height of the dyadic range tree). 
We also measured the running time of these techniques while varying the
data densities from $10^{-1}$ to $10^{-3}$. 
We observe the same trend
as in Figure \ref{fig:exp2_throughput}---sampling and filtering are
always faster than the geometric mechanism. 
For smaller density values, the gain in throughput for our 
techniques is less pronounced,
i.e., about 2 to 4 times, since all methods incur the same cost for building
the dyadic ranges.

Figure \ref{fig:exp3_dyadic2} shows the same experiment when 
the data is skewed, i.e. 
the non-zeros in the data
have a higher probability to be placed close together in the input array.
This effectively changes the sparsity of the resulting
dyadic ranges, which have more zeros in certain tree nodes than
others. 
Now filtering the dyadic
ranges improves over its geometric mechanism equivalent
since it can eliminate more noise from these ``light''
nodes in the dyadic range tree. 
Overall, we conclude that the approach of dyadic ranges (and
similarly, wavelet transforms) is compatible with the shortcut
summarization approach, and is effective when we
anticipate that queries will touch any more than a small fraction of
the data space. 

\para{Consistency checks.}
We next experiment with applying consistency checks to the 
dyadic ranges, as discussed in Section~\ref{sec:dyadic}. 
This is done for plain filtering, and on the output of filtering prior
to priority sampling. 
Note that applying these checks on the simple geometric mechanism is not
helpful, since there are very few entries set to zero which can be
used to filter descendant nodes. 
Figure~\ref{fig:exp3_dyadic_consistency} shows  
that applying consistency checks reduces the
errors of our two techniques by 30\% to 60\%. 
This confirms that consistency checks help improve accuracy when the
data is highly sparse and non-uniform, since they further eliminate
noise from originally zero entries.  
For more uniform datasets, the same trend is present, but is less
pronounced: the improvement is closer to 10\%. 

\subsection{OnTheMap Data}
We evaluate performance on the ``OnTheMap'' data
described in Appendix~\ref{app:datasets}. 
The data is actually synthesized from real
individuals but is designed to represent realistic patterns.
For the purposes of our study, 
 we treat it as the ground truth $M$ and compare our techniques for
 protecting the privacy of these (synthetic) individuals.  
We first apply filtering and sampling over this data set, and show 
results for a summary of size $3 \times 10^6$. 
Figure~\ref{fig:exp4_map1} shows that
applying filtering to this data is inaccurate, partly 
due to the high variance of the data: deleting too
many small values actually wipes out a lot of the original signal. 
Priority sampling and filter-priority sampling with a small threshold give
better accuracy: not much worse than the geometric mechanism (whose
output size is %
more than 5 times larger) for most 
queries, and actually better for large queries which cover 20\% or
more of the data space. 
On these large subset queries, priority sampling
outperforms the other techniques, but filter-priority is not far behind. In terms of relative errors, all queries for subsets covering 5\% or above of the data size have errors less than 0.8\%. 

To compare dyadic ranges, we linearize the data by sorting by the work location
id. 
The size of the published dyadic ranges used is $5 \times 10^5$, 
smaller than original data cardinality. 
Figure~\ref{fig:exp4_map2} shows the benefit of
dyadic ranges for 
even small-medium range sizes (around 1\% of the data size). 
Filtering improves the accuracy over 
just applying the geometric mechanism. This can be
explained as a case where the data has a very high variance, and most of the small-valued entries are masked with noise, 
hence we do not need to publish a large sample. We further apply consistency checks on this data. Figure \ref{fig:exp4_map_consistency}  shows that having these checks helps improve the accuracy about 10\%, similarly to the above setting where the non-zeros are placed uniformly.

We also compared to the method of 
Machanavajjhala {\em et al.} to anonymize data via synthetic data
generation \cite{Machanavajjhala:Kifer:Abowd:Gehrke:Vilhuber:08}. 
We applied their method using parameters 
$\epsilon=1$, $\delta= 10^{-4}$,
to satisfy
$(\epsilon, \delta)$-probabilistic differential privacy. 
While this approach has been shown to approximately 
preserve features such as the distribution of 
commuting distance in the data
\cite{Machanavajjhala:Kifer:Abowd:Gehrke:Vilhuber:08}, it does not
seem to help to accurately answer range queries. 
For example, 
over the same set of queries we observe absolute error 
over three times that of the geometric mechanism with the same
parameters. 
Further, it does not seem possible to find a setting of the parameters 
$\alpha(i)$ required by the method to obtain stricter privacy
guarantees (i.e. $\epsilon<1$) such as those we focus on here 
(our experiments are carried out with the stronger guarantee of
 $\epsilon=0.1$). 

\section{Concluding Remarks}
Differential privacy represents a powerful mechanism for releasing
data without compromising the privacy of the data subjects. 
We have shown that our shortcut approach is an effective way to
release data under differential privacy
 without overwhelming the data user or the data owner.
The accuracy of query answering from the released summary compares
favorably to the laborious approach of publishing vast data with
geometric noise, and can actually help the user to improve their
utility by removing a lot of the noise without compromising the
privacy. 
Both filtering and sampling are effective in different
cases, but the combined filter-priority method seems generally useful
across a wide variety of settings. 
On data which is sparse, as is the case for most
realistic examples, the cost creating the summary is low, and the
benefit only improves as the data dimensionality increases. 
When we expect that range queries are prevalent, the summaries can
be applied to techniques such as dyadic ranges. 
The benefits of the summary still hold (compact, accurate, fast to
compute) and the benefit compared to the base method is felt for
moderately short queries, corresponding to a small fraction of the
input space.

{%

}

\clearpage
\appendix
\allowdisplaybreaks
\section{Datasets}
\label{app:datasets}

\para{Census Income Data.}
Census Income Data has attributes ({\em Age, Birthplace, Occupation, Income}).
Samples can be downloaded from \url{www.ipums.org}. 
To represent in a contingency table, we considered income at the
granularity of \$10,000 multiples, and age to multiples of 5 years. 
Under various settings, we observed data density of 3\% to 5\%.

\para{OnTheMap Data.}
The US Census Bureau makes available data describing 
commuting patterns US residents, as the number of 
 people for each work-home combination (at the census block level),
 together with other information 
such as age ranges, salary ranges, and job types. 
We consider the 47 states available in version 4 of the 2008 data from
\url{http://lehdmap.did.census.gov/}.
We take the location data as the first  2 digits of the census
tracts in each county; 
so each location is identified by ``county id + 2-digit tract id''. 
There are 4001 such locations, so the size of the resulting
frequency matrix is $m=1.6 \times 10^7$. 
The number of non-zeros is $\sim 8.2 \times 10^5$, 
so the data density is $\rho = 0.051$\footnote{The data is larger
  and sparser if we include more attributes: 
  adding age, salary and industry, each of which has three values, 
 increases the data size by 27 times while dropping the density to
 below $0.01$. }. 
The mean value in each non-zero cell is approximately 150, but with
very high variance: many cells have
frequency 1 or 2 (and hence should be masked by the
addition of noise).   

\para{Adult Data.}
The UCI Adult Data from the Machine Learning repository at 
\url{http://archive.ics.uci.edu/ml/datasets/Adult}
has been widely used in data mining, and prior work on
privacy~\cite{Kifer:09}.
The full data has 14 attributes, but to avoid gridding issues, we 
projected the data on categorical attributes, i.e, {\em Workclass, Education, 
MaritalStatus, Occupation, Relationship, Race, Sex}.
This generated data with a density of $0.14\%$, and an average value
in each (non-zero) cell of 9. 

\para{Telecom Warehouse Data.}
AT\&T records measurements on the performance of devices in its
network in a data warehouse. 
These measurements include attributes 
{\em deviceId, timestamp, Val}, representing
a measurement {\em Val} of each device at a given time stamp. 
For each day, many gigabytes of data are added to the warehouse. 
For several natural bucketings of the numerical attributes, the
observed density ranges from 0.5\% to 2\%.  
Therefore, the output of reporting all differentially private counts
 is 50-200 times larger.
Generating, storing and processing data in this form would increase
the associated costs by the same factors, making this vastly too
expensive. 
Nevertheless, given the company's data protection policies, 
and the need for various internal groups to analyze this data, a
practical anonymization solution is highly desirable. 

\section{Background}
\label{app:background}

\subsection{Differential Privacy}
Differential Privacy was introduced over the course of a series of
papers in the 2000s, culminating in Dwork's paper which coined the
term \cite{Dwork:06}. 
A randomized algorithm achieves $\epsilon$ differential privacy if 
the probability of output falling in some set is at most
$\exp(\epsilon)$ times the probability of the output falling in the
same set, given input which differs in the records of at most one
individual. 

\begin{definition}[From \cite{Dwork:06}]
A randomized algorithm $K$ gives $\epsilon$-differential privacy if
for all data sets $D_1$ and $D_2$ differing on at most one element,
and all $S \subset Range(K)$: 
\[ \Pr(K(D_1) \in S) \leq \exp(\epsilon) \cdot \Pr(K(D_2) \in S) \]
\end{definition}

The choice of $\exp(\epsilon)$ makes it easy to reason about 
composing results: when two pieces of
information about a data set are released with $\epsilon_1$ and
$\epsilon_2$ differential privacy respectively, their combination is
(at most) $\epsilon_1 + \epsilon_2$ differentially private. 

There is a simple recipe for generating differentially private
numerical data: compute the true answer exactly, and then output this
value with additive noise drawn from a Laplacian distribution with
parameter $(\epsilon/\Delta_q)$, where $\Delta_q$ is the
``sensitivity'' of the query, the total influence that any individual
can have on the output \cite{Dwork:06}.  
For discrete data, the Laplacian can be replaced with a symmetric
geometric distribution \cite{Ghosh:Roughgarden:Sundararajan:09}. 
In this paper, we focus 
collections of (disjoint) count queries, where the conditions
partition the input (as in the case of contingency tables and count-cubes): 
in this case $\Delta_q=1$, 
as adding or removing one individual affects at most one value in the 
collection of counts.

Although often introduced in the context of allowing queries to be answered
interactively with privacy, differential privacy naturally applies to 
  data publication. 
Effectively, the data owner chooses a collection of representative
queries to ask, and answers them all with appropriate noise. 
This has been promoted particularly in the case of histograms or
contingency tables of data \cite{Barak:Chaudhuri:Dwork:Kale:McSherry:Talwar:07,Xiao:Wang:Gehrke:10,Li:Hay:Rastogi:Miklau:McGregor:10}.
When the data is not numeric, other approaches are possible. 
The exponential mechanism \cite{McSherry:Talwar:07} gives a generic
approach: the probability of producing a particular output (amongst a
set of possible outputs) should depend exponentially on how
closely it resembles the original input. 
Directly instantiating the exponential mechanism is costly in general,
since often a huge number of possible outputs must be considered.
However, certain special cases have been shown to be efficient.

In recent years, there has been interest in the database community in
how to incorporate differential privacy into data management. 
McSherry's PINQ system adds privacy as an overlay by automating the
addition of noise to query outputs   
\cite{McSherry:09}.
Xiao {\em et al} observe that directly applying the Laplacian
mechanism to contingency tables yields poor answers to range queries,
since the noise grows with the size of the range \cite{Xiao:Wang:Gehrke:10}.
Their solution was to add  noise in the wavelet domain, which has
the same privacy properties, but reduces the error for large range
queries due to cancellations.
Similar techniques are suggested by Hay {\em et al.}
\cite{Hay:Rastogi:Miklau:Suciu:10}.  
A more general approach is given by Li {\em et al.}
\cite{Li:Hay:Rastogi:Miklau:McGregor:10}, by considering queries in
the form of arbitrary collections of linear sums of the input data
(represented as a matrix). 
A key difference of this prior work is that it considers publishing
results which are at least as big as the underlying domain size of
the data. 
In contrast, we consider the case where the total domain cardinality of the
data in contingency table form is too large to be published in full. 

Xiao {\em et al} \cite{Xiao:Xiong:Yuan:10} publish a summary
in the form of a kd-tree of counts. 
The construction algorithm examines all cells in the contingency table
of the data: a natural question is to extend our methods to produce
similar summaries over high domain cardinality data. 
Lastly, there has been some work on synthetic data generation based on
parameters derived privately from real data sets. 
Specifically, Machanavajjhala {\em et al.} propose a technique
tailored for certain types of census data
\cite{Machanavajjhala:Kifer:Abowd:Gehrke:Vilhuber:08}. 
This approach produces
an output data size comparable to the input size
when direct application of noise mechanisms would generate a dense,
noisy output. 
The guarantees produced are that the synthetic data satisfies a weaker 
($\epsilon$, $\delta$)-guarantee, for a larger value of $\epsilon$
than we tolerate. 
Instead, we aim to produce output which is based directly on the
original microdata, and offers high levels of privacy.

\subsection{Data Reduction Techniques}
\label{sec:reduction}
The topic of data summarization is vast, generating many surveys and
books on different approaches to the problem \cite{Muthukrishnan:05,Garofalakis:Gehrke:Rastogi:02,Olken:97}. 
In this paper, we consider some key techniques.
Filters are deterministic procedures which prune away parts of the
data which are thought to contribute little to the overall query
results. 
The canonical filter is the high-pass filter: data elements with low
frequencies (below the filter threshold) are dropped, while the
elements with high frequencies (aka the `heavy hitters') are retained.

Sampling is perhaps the most commonly used data reduction technique. 
A random process determines a subset of elements from the input data
to retain, along with weights to assign each element in the sample.
Queries over the full data are then approximated by applying the same
query to the (weighted) data in the sample, and reporting the sample result.
In our setting, we have initial weights attached to the input
elements.
Over such data, threshold sampling is based on a parameter $\tau$
\cite{Duffield:Lund:Thorup:03}. 
The sample includes each element $e$ with weight $w_e$ with
probability $p_e = \min(w_e/\tau,1)$. 
If sampled, the sample's weight is assigned to be 
$\max(w_e, \tau)$. 
One can verify answers to subset sum queries (sum of elements passing
a given predicate) are unbiased, i.e. correct in expectation. 

A limitation of threshold sampling is that there is not a strong
control over the size of the resulting sample: the size is
$\sum_{e} p_e = \sum_{e} \min(w_e/\tau,1)$ in expectation, but may
vary a lot. 
Priority sampling fixes the sample size to be $s$, and offers strong
guarantees about the quality of the sample relative to all methods
which generate samples of size exactly $s$ \cite{Duffield:Lund:Thorup:07}. 
Each element is assigned a priority $P_e = w_e/r$, where $r$ is drawn
 uniformly in the range [0,1]. 
The $s$ elements with the largest $P_e$ values are retained to form
the sample.
Weights are defined based on the $(s+1)$th priority value, denoted
$\tau_s$: the adjusted weight of element $e$ is $\max(w_e, \tau_s)$. 

Lastly, sketching techniques summarize the whole data set. 
The Count sketch hashes each element $e$ to one of $w$ possible buckets as
$h(e)$,
and applies a second hash function $g(e)$ to map each element to either +1 or
-1. 
Each bucket maintains the sum of all counts $w_e$ mapped there, each multiplied
by $g(e)$. 
Given a target element $e$, the count of the bucket $h(e)$ multiplied
by $g(e)$ is an unbiased estimate for $w_e$ with bounded variance. 
Taking the mean or median of $d$ independent repetitions of the
sketching reduces the variance further \cite{Charikar:Chen:Farach-Colton:02}. 
Section \ref{sec:sketch} analyzes sketches under differential privacy.

\section{Proofs and Analysis}
\label{app:proofs}

\subsection{High-pass filter}

\begin{proof}[Proof of Theorem \ref{thm:highpass}]
Clearly, {\sc Filter} has the same action as the
laborious approach
and hence has the same distribution, on the $n$ non-zero entries of $M$. 
We therefore focus on the distribution of entries which are zero in
$M$ that are represented in $M''$. For entries $i$ such that $M(i)=0$,
the probability they pass the filter is:
\begin{eqnarray*}
\Pr[M'(i) \geq \theta] &=& \sum_{x \geq \theta}  \frac{1-\alpha}{1+\alpha} \alpha^{|x|} \\
 &=& \frac{1-\alpha}{1+\alpha} \alpha^{\theta} \sum_{x \geq 0}  \alpha^{x} \\
 &=& \frac{1-\alpha}{1+\alpha} \alpha^{\theta} \frac{1}{1-\alpha} \\
 &=& \frac{\alpha^\theta}{1+\alpha}   \triangleq p_\theta 
\end{eqnarray*}

\noindent
Hence, the number of zero entries upgraded, i.e., which
pass the filter after noise is added, follows a binomial distribution
$\bin(m-n,p_\theta)$.  
We need to select this many locations $i$ in $M$ such that
$M(i)=0$. 
Assuming $m\gg n$, we can just perform rejection sampling: 
  randomly pick a location uniformly from the $m$ possibilities, and
  accept it if $M(i)=0$; otherwise,  reject and repeat. 
The distribution of  values which were zero but whose noise exceeds
$\theta$ is:
\begin{eqnarray*}
\Pr[ M'(i) = b | M'(i) \geq \theta] &=& \frac{\Pr[ M'(i) = b]}{\Pr[M'(i) \geq \theta]} \\
 &=& (1-\alpha) \alpha^{b - \theta} 
\end{eqnarray*}
The CDF of this distribution is:
\begin{eqnarray*}
 \Pr[X \leq  x] &=& \sum_{M'(i) \leq x} \Pr[M'(i) | M'(i) \geq \theta] \\
 &=& \sum_{b=\theta}^x (1-\alpha) \alpha^{b-\theta} \\
 &=& 1 - \alpha^{x- \theta +1},  \forall x \geq \theta  
\end{eqnarray*}
Hence the algorithm which uses this distribution operates as claimed. 
\end{proof}

Given this form of the CDF, it is straightforward to draw from it: we
draw a uniform random value $r$ in $(0,1)$, 
and invert the equation to find for
which $x$ is $\Pr[X\leq x] = r$.
This value of $x$ is then distributed according to the required
distribution, and can be used as the value for $M''(i)$.

The two-sided filter, which passes values $|M'(i)|\geq \theta$, is
very similar. 
The only differences are that $p_\theta$ is now equal to 
$\frac{2}{(1+\alpha)}\alpha^\theta$, and the CDF now includes both
positive and negative values.
The new CDF  sets $\Pr[|X| \leq x] = 1-\alpha^{x-\theta+1}$. 
So we draw from the same distribution as before, but now flip a fair
coin to determine the sign of $M'(i)$.

\para{Time and Space Analysis.}
The running time of this algorithm is composed of the time to process
the $n$ non-zero entries, and the time to upgrade the
$k$ zero entries. 
Since adding noise and drawing a value for each upgraded zero both
take constant time, the overall cost is $O(n+k)$. 
The value $k$ depends on the parameter $p$: in expectation,
$\E[k]=np_\theta$, which in turn depends on $\theta$.

Suppose we have a target output size of $t$ tuples (chosen, perhaps,
roughly proportional to the original data size $n$). 
How should we choose $\theta$?
First, we focus on the zeros: if we pick 
$p_\theta \approx t/(m-n)$, then we expect to have around $t$ zeros
upgraded. 
This leads us to pick 
\[\theta= \frac{\log(\frac{(1+\alpha)t}{m-n})}{\log \alpha}, \] for one-sided filters.
We can apply this threshold, and consider the output of the 
{\sc Filter} algorithm: if it is sufficiently close to $t$, we can
accept the output. 
Otherwise, if the output is too large we need to choose a higher $\theta$.
Rather than re-running the algorithm, we can simply take its output
and apply a higher filter $\theta'$ on it until we reach the desired
sample size.
If the output of the algorithm is too small, we can re-run the
{\sc Filter} algorithm with the same $\theta$ value
 until we obtain a sufficiently large summary. 

\noindent
{\bf Query Answering.}
Naively,  the filtered output can be used directly to answer queries. 
This is somewhat biased, in that all small values in $M'$ have been
replaced with zero. 
However, it may work well in practice if $\theta$ is chosen so that
values below $\theta$ are mostly noise, while most of the original
data values in $M$ were comfortably above $\theta$. 
One can also consider other heuristic corrections to the output, such
as assuming all  values absent from $M''$ are, e.g., $\theta/2$ or some other
calibrated value. 
The subsequent methods we consider avoid this issue by providing
unbiased estimators for query answering.

\subsection{Threshold sampling}
\label{app:threshold}
\begin{proof}[Proof of Theorem \ref{thm:threshold}]
The non-zero entries are sampled with the probabilities defined for
threshold sampling, so certainly both approaches have the same
distribution on the $n$ non-zero entries of $M$. 

Consider the zero entries. 
For every zero entry $i$, let the geometric noise be $\nu(i)$, and it
is sampled with probability $p_i = \min(\frac{| \nu(i)|}{ \tau},1)$. 
That is:
\[ \Pr[ i \in S | M'(i) = \nu] = \min (\frac{|\nu|}{\tau}, 1) \]

The following two summations follow by standard algebra:
\begin{eqnarray}
\sum_{x=0}^v \alpha^x  &=&  \frac{1 - \alpha^{v+1}}{1 - \alpha} 
\label{eq:1}
\\
\sum_{x=0}^v x \alpha^x &=& \frac{\alpha}{(1-\alpha)^2} (1 - (v+1) \alpha^v + v \alpha^{v+1}) 
\label{eq:2}
\end{eqnarray}

Using equations \eqref{eq:1} and \eqref{eq:2}, 
the probability that $i$ is in the sample is given by:
\begin{align*}
\Pr[i \in S ] &= \sum_\nu \Pr[ i \in S | M'(i) = \nu] \Pr( M'(i) = \nu)  \\
	=& \sum_{|\nu| \leq \tau} \frac{|\nu|}{\tau}
        \frac{1-\alpha}{1+\alpha} \alpha^{|\nu|} + \sum_{|\nu| > \tau}
        \frac{1-\alpha}{1+\alpha} \alpha^{|\nu|}  \\
	=& 2 \left (\frac{1- \alpha}{\tau(1+\alpha)} \sum_{\nu
          =0}^\tau  \nu \alpha^\nu +  \sum_{\nu > \tau}
        \frac{1-\alpha}{1+\alpha} \alpha^{\nu} \right ) 	\\ 
	=&  2 \left ( \frac{\alpha}{\tau(1- \alpha^2)} (1 - (\tau+1)
        \alpha^{\tau} + \tau \alpha^{\tau+1} )  +
        \frac{\alpha^{\tau+1}}{1+ \alpha} \right )  \\ 
	=& \frac{2\alpha (1 - \alpha^\tau)}{\tau(1- \alpha^2)}  
   \triangleq    p_\tau
\end{align*}

Since each zero has the same, equal chance of being sampled by the
threshold sampling, the number sampled follows a binomial
distribution $\bin(m-n, p_\tau)$.

 Given that $i$ is picked for the sample, its value $M'(i)$ is
 conditioned on the fact that it was sampled:
\begin{eqnarray*}
\Pr[M'(i) = \nu | i \in S] &=& \frac{\Pr[i \in S| M'(i) = \nu] \Pr[M'(i) = \nu]}{ \Pr[(i) \in S]} \\
	&=& \frac{ \min( \frac{|\nu|}{\tau}, 1) \cdot  \frac{1-\alpha}{1+\alpha} \alpha^{|\nu|}}{p_\tau} \\
\end{eqnarray*}
\begin{align*}
\text{If $|\nu| > \tau$,} \quad 
\Pr[M'(i) = \nu | i \in S]  &= \frac{\frac{1-\alpha}{1+\alpha}
  \alpha^{|\nu|}}{\frac{2\alpha(1 - \alpha^\tau)}{\tau(1-\alpha^2)}} \\ 
	&= \frac{\tau(1-\alpha)^2 \alpha^{|\nu|}}{2\alpha (1-\alpha^\tau)}
\end{align*}

\noindent

\begin{align*}
\text{ If $|\nu| \leq \tau$,} \quad
\Pr[M'(i) = \nu | i \in S]  = \frac{(1-\alpha)^2 |\nu|
  \alpha^{|\nu|}}{2\alpha (1-\alpha^\tau)}
\end{align*}

\noindent Let $C_\tau= \frac{1 } {2\alpha ( 1 - \alpha^\tau) }$; 
then the CDF of this distribution, at any $\nu$, can be computed by summing the probability of $\Pr[M'(i)=x| i \in S]$, $\forall x \leq \nu$, given by the above two expressions. It can be broken into four pieces and
  written as follows.

\[
\begin{array}{ll}
 \Pr[X = \nu \leq -\tau] &=   \tau \alpha^{-\nu}C_\tau(1-\alpha)  \\[6pt]
  \Pr[-\tau < X = \nu \leq 0] &= C_\tau ( \tau \alpha^\tau (1-\alpha) -\nu
  \alpha^{-\nu} + (\nu + 1) \alpha^{-\nu+1} \\
  & \qquad \qquad - \tau \alpha^\tau + (\tau-1)\alpha^{\tau+1}) \\
    &= C_\tau (- \nu \alpha^{-\nu}  + (\nu +1) \alpha^{-\nu +1} -
  \alpha^{\tau+1}) 
\\[6pt]
\Pr[ 0 < X = \nu \leq \tau] 
  &=  \frac{1}{2} +  \alpha C_\tau(1 - (\nu+1)\alpha^{\nu} + \nu\alpha^{\nu+1})  
  \\[6pt]
\Pr[\tau < X = \nu]  &=  
   \frac{1}{2} + C_\tau (\alpha (1 - (\tau+1)\alpha^{\tau} + \tau \alpha^{\tau+1})\\ 
    & \qquad \qquad + \tau\alpha^{\tau+1} (1 - \alpha^{\nu-\tau})(1-\alpha) ) \\
  &= \frac{1}{2} + \alpha C_\tau(1 - \alpha^\tau - \tau\alpha^\nu (1-\alpha))
\end{array}\]

Thus choosing $M''(i) = \nu$ according to this CDF has the correct
distribution. 
\end{proof}

\para{Time and Space Analysis.}
It follows immediately from the description of the algorithm that the
sample can be generated in time $O(n + k) = O(n + p_\tau (m-n))$. 
We now analyze how to choose $\tau$ given a desired sample size $t$. 
We can do the same as for estimating $\theta$ in high-pass filters, i.e., choosing $\tau$ so
that the number of zeros sampled is about $t$. Or, we can also make a more accurate estimation by taking 
into account the statistics of the non-zeros as follows.
The expected size of the threshold sample is
$t = \sum_{i} \min(|M'(i)|/\tau, 1)$. 
On high cardinality data, where $\tau$ will be large so that the
sample size is bounded, we may approximate this by 
\[\sum_{i} \frac{|M'(i)|}{\tau} \leq \frac1\tau \sum_{i} (|M(i)| + |G(\alpha)|)\]
where $G(\alpha)$ is the geometric distribution with parameter
$\alpha$.
Using \eqref{eq:2}, we have that
\[\E[|G(\alpha)|] = 2\sum_{j=0}^{\infty} \frac{1-\alpha}{1+\alpha} j
\alpha^j = \frac{1-\alpha}{1+\alpha} \frac{2\alpha}{(1-\alpha)^2}
= \frac{2\alpha}{1-\alpha^2}. \]
Hence, in expectation this approximation is
\[\|M\|_1/\tau + m\E[ |G(\alpha)|]/\tau = 
  (\|M\|_1 + 2m\alpha/(1-\alpha^2))/\tau\]
\noindent
Rearranging, this suggests that we should pick $\tau$ in the region of 
$\frac1t(\|M\|_1 + 2m\alpha/(1-\alpha^2))$ for a desired sample size $t$. 
As with the high-pass filter, we can draw a sample based on an
estimate of $t$, then refine our choice of $\tau$ to achieve a desired
sample size.  
However, this is taken care of more directly when we build a priority
sample of fixed size, discussed in the next section.

\para{Query Answering.}
Each element is sampled with probability $p_i = \min(|M'(i)|/\tau,1)$.
If we scale each sampled item's 
weight by this probability, we obtain an unbiased
Horvitz-Thompson estimator for the weight \cite{Duffield:Lund:Thorup:03}. 
The data owner can scale all values by this weight before release. 
This allows queries to be answered by evaluating the query over the
sample, using the adjusted weights. 
However, for very small queries (e.g. point queries), it can more
accurate to use the (biased) estimator of the unadjusted weight.  

\subsection{Priority sampling}

Theorem \ref{thm:priority} follows immediately from the previous
discussion, so we omit a formal proof for brevity.

\noindent
{\bf Query Answering.}
As with threshold sampling, there is an unbiased way to adjust the
weights: 
letting $\tau_s$ be the $(s+1)$th priority, 
we assign each item a weight with magnitude $\max(\tau_s,|M''(i)|)$ while keeping the sign of $M''(i)$.
This is unbiased, and has low variance for arbitrary subset queries
compared to any other possible sampling scheme which builds a sample
of $s$ items \cite{Duffield:Lund:Thorup:07}. 

\subsection{Threshold sampling with High-pass filter}
\label{app:samplefilter}

We have two parameters: $\theta$ for the filter, and $\tau$ for the
sampling. 
If we set $\tau\leq \theta$, then every item which passes the filter enters the
sample, i.e. we have just filtering. 
If we set $\theta=0$, then every item passes the filter, i.e. we have
just sampling. 
The interesting range occurs when we have $0 < \theta < \tau$: in this
regime, items in $M'$ with absolute weights below $\theta$ are suppressed, above
$\tau$ are guaranteed to be included in $M''$, and in between have a
chance proportional to their weight to be included. 

As in previous cases, we can handle the non-zero elements of $M$
directly: we add noise drawn from the appropriate geometric
distribution to obtain $M'(i)$, and first filter then sample the
element. 
It is the large number of zero elements which we must treat
more carefully to keep the process efficient. 
We can follow the analysis in previous sections to help us determine how
to proceed. 

For any zero entry, the distribution of its noisy value after
threshold filtering, $M'_\theta(i)$ is: 
$\forall |\nu| \geq \theta$,
\begin{eqnarray*}
\Pr[M'_\theta (i) = \nu] & = &\frac{1-\alpha}{1+\alpha}\alpha^{|\nu|} \\[6pt]
\Pr[M'_\theta(i) = 0 ] &=& \Pr[M'(i) < \theta] 
= \sum_{|x| < \theta} \frac{1-\alpha}{1+ \alpha} \alpha^{|x|} \\
 &=& \frac{1-\alpha}{1+\alpha} 2 \sum_{x< \theta} \alpha^x 
 = 1 - \frac{2\alpha^\theta}{1+\alpha}
\end{eqnarray*}

Given the parameter $\tau$, the value of $p_{\theta,\tau}$, the probability of the
element being sampled, is:
\begin{eqnarray*}
p_{\theta,\tau} &=& \sum_\nu \Pr[ i \in S | M_\theta'(i) = \nu] \Pr[ M_\theta'(i) = \nu] \\
   &=& \sum_{|\nu| > \tau} \frac{1-\alpha}{1+\alpha}
   \alpha^{|\nu|} + \sum_{\theta \leq |\nu| \leq \tau}
   \frac{1-\alpha}{1+\alpha} \alpha^{|\nu|} \frac{|\nu|}{\tau} \\ 
   &=& 2 \left ( \sum_{\nu > \tau} \frac{1-\alpha}{1+\alpha}
   \alpha^{\nu} + \sum_{\theta \leq \nu \leq \tau}
   \frac{1-\alpha}{1+\alpha} \alpha^{\nu} \frac{\nu}{\tau} \right ) \\ 
 &=& 2 \left ( \frac{\alpha^{\tau+1}}{1+\alpha} +
   \frac{\alpha}{\tau(1-\alpha^2)}(\theta \alpha^{\theta-1} -
   (\theta-1)\alpha^\theta \right.
   \left. - (\tau+1) \alpha^\tau  + \tau \alpha^{\tau+1}) \right ) \\
 &=& \frac{2}{\tau(1-\alpha^2)} ( \theta \alpha^{\theta} - (\theta-1) \alpha^{\theta+1} - \alpha^{\tau+1})
\end{eqnarray*}
Observe that for the case $\theta=0$, this simplifies to 
$\frac{2\alpha(1-\alpha^\tau)}{\tau(1-\alpha^2)}$, the expression for
$p_\tau$ from the previous section. 
As before, we can use $p_{\theta,\tau}$ to determine the number of zero locations
from $M$ to sample.
Given such a location, conditioned on it being included in the sample,
the distribution of its value is now
\begin{eqnarray*}
\Pr[ M_\theta'(i) = v | i \in S] &=& \frac{\Pr[i \in S| M_\theta'(i) = v] \Pr[M_\theta'(i) = v]}{\Pr[ i \in S]} \\
 &=& \tau C_{\theta,\tau}  (1-\alpha)^2 \alpha^{|v|} \min (\frac{|\nu|}{\tau}, 1)
\end{eqnarray*}
where the constant 
$C_{\theta,\tau}= (2( \theta \alpha^{\theta} - (\theta-1) \alpha^{\theta+1} - \alpha^{\tau+1}))^{-1}$.

\smallskip
\noindent
The CDF of the sampled values has four cases, $\Pr[ X \leq
  \nu] = $
\[
\begin{array}{ll}
\tau C_{\tau,\theta}   (1-\alpha)  \alpha^{-\nu}, 
     & \mbox{if } \nu \leq -\tau  \\
   C_{\tau,\theta}  (-\nu \alpha^{-\nu} + (\nu+1)\alpha^{-\nu+1} - \alpha^{\tau+1}), 
  & \mbox{if }  -\tau < \nu \leq -\theta \\ 
   \frac{1}{2} + C_{\tau,\theta} (\theta \alpha^{\theta} - (\theta-1) \alpha^{\theta+1} \\
  \qquad \qquad \qquad 
    - (\nu+1)\alpha^{\nu+1} + \nu \alpha^{\nu+2} ), 
     & \mbox{if } \theta \leq \nu \leq \tau \\
   1- \tau C_{\tau,\theta}  (1-\alpha) \alpha^{v+1}  , & \mbox{if } \nu  > \tau
\end{array}
 \]

Since we can answer threshold samples on the filtered data,
  we can also support priority samples via the observation of 
Section \ref{sec:priority}, that we can extract a priority sample from
a threshold sample.

\para{Query Answering.}
The standard sampling estimation schemes allow us to provide unbiased
estimates over the underlying data.
In this case, the estimates are of the filtered version of $M'$. 
Since filtering with a low threshold removes much of the noise, being
unbiased over the filtered $M'$ may be a good approximation of the
original $M$ data, while remaining private. 

\subsection{Range Queries}

\begin{proof}[Proof of Theorem \ref{thm:dyadic}]
The approach is quite direct: from $M$, we produce the (exact) transform
into either wavelets or dyadic ranges, then apply the above algorithms
to this transform. 
The theorem follows by observing that the total number of non-zero
values in the transform of $M$ can be bounded in terms of $n$ and
$m$. 
We outline the case for dyadic ranges; wavelets are similar. 

Consider each non-zero in $M$ as a leaf in a binary tree.  
The number of non-zeros in the dyadic range transform of $M$ is the
number of distinct nodes touched by following the paths from each non-zero
to the root. 
Each non-zero can touch at most one node in any level, giving a crude
bound of $n \log m$. 
This can be tightened by observing that for levels close to the root,
there may be fewer than $n$ nodes. 
More precisely, for the top $\log n$ levels, there are at most $n$
nodes in total.
Thus we  bound the total number of nodes touched, and hence the
number of non-zeros in the transform as:
\smallskip
\hspace*{6mm} { $O(n + n(\log m - \log n)) = O(n \log (m/n))$. }
\end{proof}

\subsection{Sketch Summarization}
\label{sec:sketch}
Both filtering and sampling keep information about a selected subset
of locations in the original data space, and drop information about
the remaining locations. 
In contrast, sketch techniques bring together information about every
point in the original data. 
A first approach is to generate the noisy data $M'$, and produce
the sketch of this: using the Count Sketch (described in Section
\ref{sec:reduction}), we would have significant noise in each bucket
in the sketch. 
We can improve on this considerably by observing that, viewing the
sketch as a query, the sketch has low sensitivity. 
That is, we can view sketching as a mechanism for publishing data
similar to histograms: once we fix the hash function $h$, each bucket
contains the sum of counts of a set of individuals, and this forms a
partition of the input data. 
Thus, over $d$ rows of the sketch, the sensitivity 
is $\Delta_s = 2d$, and we can achieve privacy by adding the
appropriate amount of random noise to each entry in the sketch.
Consequently:

\begin{lemma}
The sketch mechanism generates a sketch of size $w \times d$ 
in time $O((n+w)d)$. 
The variance of point estimates from the sketch is 
$O(\|M\|_2/wd + d/\epsilon^2)$. 
\end{lemma}

The time cost follows from the fact that we have to map each non-zero
location in the input to $d$ rows of the sketch, and then add noise to
each entry.  
The variance of each estimate, due to sketching, is
proportional to the Euclidean norm of the data scaled by the number of
buckets, $\|M\|_2/w$
\cite{Charikar:Chen:Farach-Colton:02}. 
The noise added for privacy simply adds to this variance, in the
amount of $O(d^2/\epsilon^2)$ for noise with parameter $(\epsilon/d)$. 
Taking the average of $d$ repetitions scales the variance down by a
factor of $d$ to 
$O(\|M\|_2/wd + d/\epsilon^2)$. 

Thus there is a tradeoff for setting the parameter $d$:
increasing $d$ reduces the sketching error, but increases the privacy
error.  
In an experimental study, we able to use large values of $w$, so the
second term dominated, meaning that the optimal setting was to pick
small values of $d$, such as $d=1$. 
We observed that the error was
considerably higher than for sampling/filtering, 
so we omit a detailed study of sketching from this presentation. 

\end{document}